\documentclass[aps,prb,twocolumn,showpacs,amsmath,amssymb,amsthm,superscriptaddress,groupedaddress]{revtex4}
\usepackage{graphics}
\usepackage{epsfig}
\usepackage[colorlinks=true,citecolor=magenta,linkcolor=red,linktocpage=true,pagebackref=false]{hyperref}
\usepackage{soul} 
\usepackage[usenames, dvipsnames]{color}
\usepackage{braket}

\begin{document}
\title{
Auto- versus cross-correlation noise 
}
\author{Michael Moskalets}
\email{michael.moskalets@gmail.com}
\affiliation{Department of Metal and Semiconductor Physics, NTU ``Kharkiv Polytechnic Institute", 61002 Kharkiv, Ukraine}

\date\today
\begin{abstract}
Expressing currents and their fluctuations at the terminals of a  multi-probe conductor in terms of the wave functions of carriers injected into the Fermi sea  provides new insight into the physics of electric currents. 
This approach helps us to identify two physically different contributions to shot noise.
In the quantum coherent regime, when current is carried by non-overlapping wave-packets, the product of current fluctuations in different leads, the cross-correlation noise, is determined solely by the duration of the wave packet. 
In contrast, the square of the current fluctuations in one lead, the auto-correlation noise, is additionally determined by the coherence of the wave-packet, which is associated with the spread of the wave packet in energy. 
The two contributions can be addressed separately in the weak back-scattering regime, when the auto-correlation noise depends only on the coherence. 
Analysis of shot noise in terms of these contributions allows us, in particular, to predict that no individual travelling particles with a real  wave function, such as Majorana fermions, can be created in the Fermi sea in a clean manner, that is, without accompanying electron-hole pairs.
\end{abstract}
\maketitle


\section{Introduction}
\label{intro}

Recently the quantum tomography of a single electron wave function was demonstrated experimentally.\cite{{Jullien:2014ii},{Bisognin:2020dk}} 
In both experiments, a periodic stream of single-electron wave packets was mixed with a low amplitude electrical probe signal at the electron wave splitter, a quantum point contact and the resulting electrical noise averaged over long time was measured. 
However, if the fluctuations of an electrical current within one output lead were measured in one experiment, the correlations of currents flowing within both output leads were measured in another.    
In a sense, these works are the culmination of a number of recent works where the cross-correlation noise was measured to count electrons emitted per period \cite{Bocquillon:2012if,Dubois:2013ul}, to demonstrate a tunable fermionic anti-bunching \cite{Dubois:2013ul,Bocquillon:2013dp,Marguerite:2016ur,Glattli:2016wl}, and the auto-correlation noise at high \cite{Mahe:2010cp,Parmentier:2012ed} and low \cite{Maire:2008hx,Gabelli:2016uz} frequencies was measured to identify a single-electron emission regime.  
These experimental advances stimulate us to take a closer look at how exactly the quantum properties of a wave packet manifest themselves in  the measured electrical noise \cite{Landauer:1998jh,Reznikov:1998kn,Blanter:2000wi}. 

Note also that tomography of the density profile of solitary electrons was successfully realized in Ref.~\onlinecite{Fletcher:2019jc} using the  measurements of the electrical current rather than noise.  

\begin{figure}[t]
\centering
\resizebox{0.99\columnwidth}{!}{\includegraphics{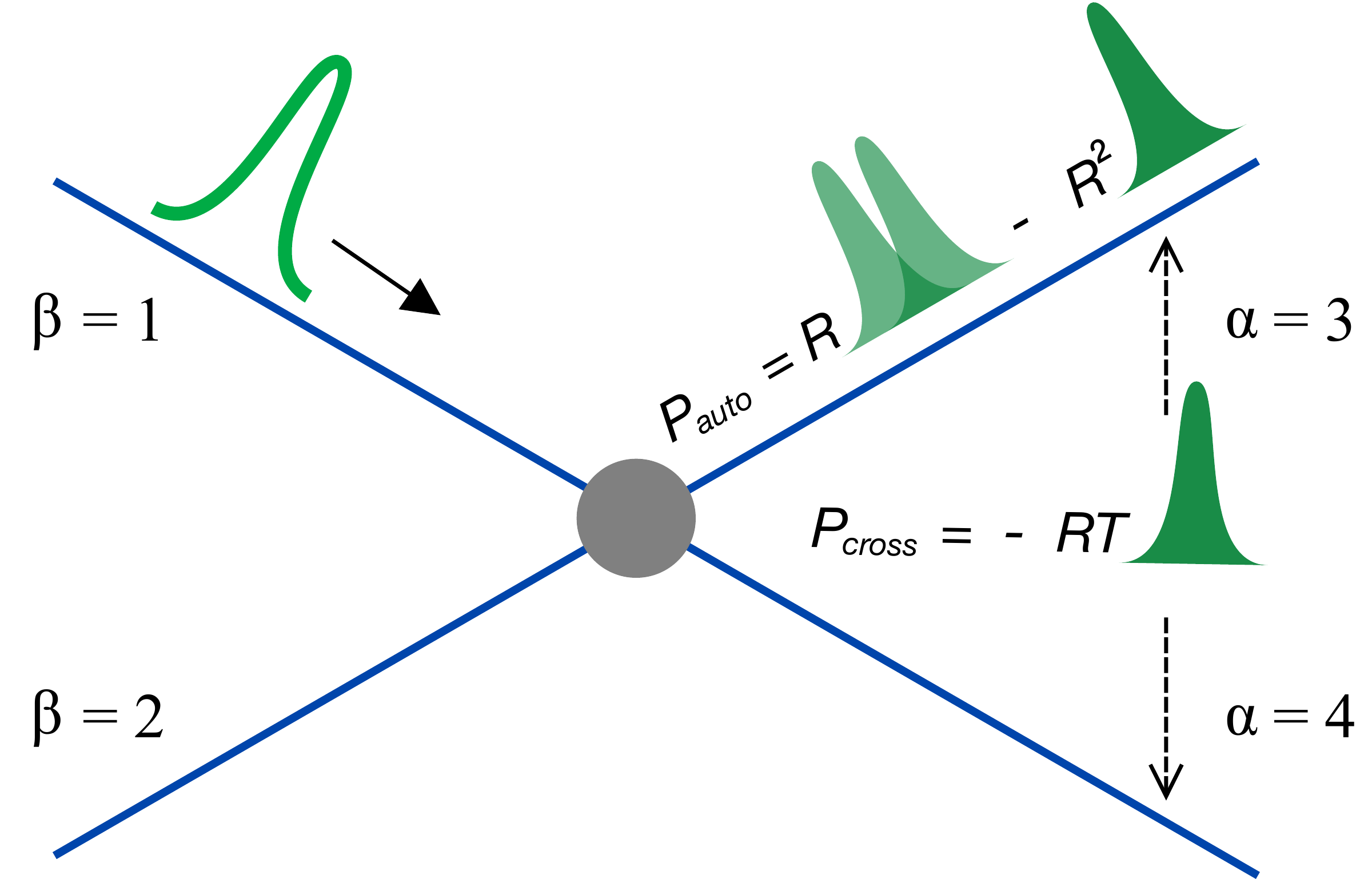} }
\caption{
Shot noise, see Eq.~(\ref{n01}), of a single-electron wave packet injected on top of the Fermi sea consists of two parts. 
One is determined by the density profile of the wave packet, shown as a filled hump, and the other is determined by both the coherence of the wave packet,  shown as a double hump, and the Fermi sea, shown as a blue line. 
While the former contributes to both auto-correlation noise, ${\cal P}_{auto}$, and cross-correlation noise, ${\cal P}_{cross}$, the latter contributes to ${\cal P}_{auto}$ only. 
The solid arrow indicates the wave packet in incoming channel $ \beta = 1$, shown as an empty  hump. 
The filled circle represents the wave spitter with reflection probability $R$ and transmission probability $T=1-R$. 
The two dashed arrows point to the two outgoing leads $ \alpha = 3$ and $ \alpha = 4$.}
\label{fig1}
\end{figure}

Here, in contrast to previous works \cite{Albert:2010co,Grenier:2011js,Grenier:2011dv,Dubois:2013fs,Grenier:2013gg,Ferraro:2013bt,Thibierge:2015up,Gaury:2016ez,Rech:2017be,Roussel:2017hu,Ferraro:2018um,Misiorny:2017ua,Glattli:2017vp,Dittmann:2018vea,Rebora:2020jd,Yue:2020uo,Roussel:2020vd}, -- for a review, see Ref.~\onlinecite{Bauerle:2018ct}--  I will focus on comparing auto- and cross-correlation noise. 
I will show that, in the case of a periodic train of non-overlapping single-electron wave packets  scattered off the wave splitter with reflection probability $R$, there are two  contributions to noise. 
They can be addressed separately by measuring both tapes of noise in the weak back-scattering regime, $R \ll 1$.   
In particular, for a single-electron wave packet $ \Psi\left(  t \right) = e^{ - \frac{ i }{ \hbar  } \mu  t  } \psi \left(  t \right)$ injected during one period into a chiral Fermi sea  and reflected into a detector at a wave splitter, see Fig.~\ref{fig1},  the auto-correlation, ${\cal P}_{auto}^{ex}$, and cross-correlation, ${\cal P}_{cross}^{}$, noise power at frequency $ \omega$ and at zero temperature are calculated as follows,

\begin{eqnarray}
{\cal P}_{auto}^{ex}\left(  \omega \right )  &=&  
R \frac{ e^{2} }{ {\cal T} _{0} }  
\int\limits_{- \infty}^{ \infty } d \tau 
e^{i \omega \tau }
\frac{ {\rm Im} \, {\cal C}\left(   \tau \right) }{  \pi \tau  }
,
\label{01} \\  
{\cal P}_{cross}^{}\left(  \omega \right)  &=& 
-R \frac{ e^{2} }{ {\cal T} _{0} }   
\left | {\cal N}\left(   \omega \right) \right |^{2}
, 
\nonumber 
\end{eqnarray} 
\noindent \\
where the superscript ${ex}$ indicates the excess over the equilibrium value,   $e$ is an electron charge, $ {\cal T} _{0}$ is a period,  
${\cal C}\left(   \tau \right)$ is the integrated over time coherence of the wave function envelope,\cite{Haack:2011em,Haack:2013ch} 

\begin{eqnarray}
{\cal C}\left(   \tau \right) &=& \int\limits_{- \infty  }^{ \infty  }  dt  \psi^{*}\left(  t+ \tau \right)\psi\left(  t \right) ,
\label{01-1}
\end{eqnarray}
\noindent \\
and ${\cal N}\left(   \omega \right)$ is the Fourier transform of the wave packet density,

\begin{eqnarray}
{\cal N}\left(   \omega \right) &=& 
\int\limits_{- \infty}^{ \infty } d \tau e^{i \omega \tau }\left | \psi^{}\left(  \tau \right) \right |^{2}  .
\label{01-2}
\end{eqnarray}
\noindent \\
Thus, we see that in the weak backscattering regime, the auto-correlation noise and the cross-correlation noise are determined by essentially different quantities. 
One can say that they provides somehow complementary information.  
The cross-correlation noise is sensitive merely to the shape of a wave packet, that is, to its duration in time. 
While the auto-correlation noise is rather sensitive to how different parts of the wave packet correlate with each other. 
Such correlations are related to the spread of the wave packet in energy,  
which is supported by the reasoning of Ref.~\onlinecite{Parmentier:2012ed} that only those particles whose energy exceeds the Fermi energy by more than $\hbar \omega$ contribute to the auto-correlation noise at the frequency $ \omega$.  
The following examples provide additional evidence of the duality of the information provided by both types of noise. 

A Lorentzian voltage pulse, one per period, $eV(t) = 2 \hbar \Gamma _{\tau}  \left(  t^{2} + \Gamma _{\tau}^{2} \right)^{-1}$,  applied to the Fermi sea with energy $ \mu$ and at zero temperature creates a single-electron wave packet \cite{Levitov:1996,Ivanov:1997}, named a leviton \cite{Dubois:2013ul}. 
This excitation should be understood in such a way that the voltage pulse shakes the Fermi sea and excites just a single electron on its surface.\cite{Flindt:13}  
The wave function of a leviton, $\Psi^{L}(t) = e^{ - \frac{ i }{ \hbar  } \mu  t  }\psi^{L}(t)$, has the following envelope function, \cite{Keeling:2006hq}

\begin{eqnarray}
\psi^{L}(t) = 
\frac{  1 }{  \sqrt{\pi  \Gamma _{\tau}  } } 
\frac{ \Gamma _{\tau} }{ t - i \Gamma _{\tau} } .
\label{02}
\end{eqnarray}
\noindent \\
I assume $ \Gamma _{\tau} \ll {\cal T} _{0}$ to avoid overlap between the successive wave packets.  
Using the above equation in Eq.~(\ref{01}), I calculate,  

\begin{eqnarray}
{\cal P}_{auto}^{ex,L}\left(  \omega \right) = - {\cal P}_{cross}^{L}\left(  \omega \right) &=& R 
\frac{ e^{2} }{ {\cal T} _{0} }  
e^{- \left | \omega \right |  2 \Gamma _{\tau}}
.
\label{03}  
\end{eqnarray} 
\noindent \\
The fact that the auto- and cross-correlation noise show the same frequency dependence is due to the fact that there is only one time parameter in the problem, $ \Gamma _{\tau}$, which defines both the characteristic energy and its mean fluctuations, $ \hbar/ \left(  2 \Gamma _{\tau} \right)$, \cite{Keeling:2006hq} and the characteristic width in time, $2 \Gamma _{\tau}$.

In the next example the shape and the energy distribution are not related so tight and ${\cal P}_{auto}^{ex}\left(  \omega \right)$ and $-{\cal P}_{cross}^{}\left(  \omega \right)$ become different.

Let us consider a quantum level of half-width $ \delta$ filled with one electron and tunnel-coupled to a one dimensional Fermi sea at zero temperature. 
The energy of a level raises at a constant rapidity $c$, and crosses the Fermi level at $t=0$ when an electron is injected into the Fermi sea. 
Such regime of injection can be realized using the quantum capacitor. \cite{Buttiker:1993wh,Gabelli:2006eg,Feve:2007jx}
The wave function of the injected electron was calculated in Ref.~\onlinecite{Keeling:2008ft}, $\Psi^{c}(t) = e^{ - \frac{ i }{ \hbar  } \mu  t  }\psi^{c}(t)$, with 

\begin{eqnarray}
\psi ^{(c)}\left( t \right) &=& 
\frac{  1 }{  \sqrt{\pi  \Gamma _{\tau}  } } 
\int\limits_{0}^{ \infty} dx
e^{ - x } 
e^{  - i x \frac{ t }{  \Gamma _{\tau}  }    } 
e^{ i x^{2} \frac{ \tau_{D} }{ \Gamma _{\tau}  }  } ,  
\label{04}
\end{eqnarray}
\noindent \\
where $ \Gamma _{\tau} = \delta / c$ is the crossing time, the time it takes for a raising widened quantum level to cross the Fermi level,  $ \tau_{D} = \hbar/\left(  2 \delta \right)$ is the dwell time, an average time spent an electron on a quantum level before escaping to the Fermi sea provided that such an escape is possible, that is, after the quantum level has risen above the Fermi level. 
Note that if $ \tau_{D} \ll \Gamma _{\tau}$, then $ \psi^{c}$, Eq.~(\ref{04}), is essentially $ \psi^{L}$, Eq.~(\ref{02}). 
Notice, to get a stream of electrons we need a set of levels. 
Subsequent crossings occur with a delay of $ {\cal T} _{0} \gg \Gamma _{\tau}, \tau_{D}$.
Now equations (\ref{01}) and (\ref{04}) give us, 

\begin{eqnarray}
{\cal P}_{auto}^{ex, c}(\omega) &=& 
R \frac{ e^{2} }{ {\cal T} _{0} }  
 e^{- \left | \omega \right | 2 \Gamma _{\tau}} ,
\label{05} \\
{\cal P}_{cross}^{c}(\omega) &=& - 
R  \frac{ e^{2} }{ {\cal T} _{0} }  
\frac{ e^{-  \left | \omega \right |  2 \Gamma _{\tau}} }{ 1 + \left( \omega  \tau_{D} \right)^{2} }  .
\nonumber 
\end{eqnarray}
\noindent \\
For this model, the crossing time $2 \Gamma _{\tau}$ is the only parameter that determines  the exponential energy distribution,\cite{} the same as for the source of levitons.\cite{Moskalets:2016fm} 
This is why the auto-correlation noise is the same as in the first example. 

However, the shape of the wave packet is different from that of a leviton. 
Namely, if the dwell time is comparable with the crossing time, $ \tau_{D} \gtrsim \Gamma _{\tau}$, the density profile becomes larger then $2 \Gamma _{\tau}$, asymmetric, and with some wavy structure developing at later times. 
All this leads to additional suppression of ${\cal N}\left(  \omega \right)$ and ${\cal P}_{cross}^{c}(\omega)$ with increasing frequency. 
In the case when $ \tau_{D} \gg \Gamma _{\tau}$, the dwell time, not the crossing time determines how the cross-correlation noise decreases with frequency. 
The dwell time does not affect ${\cal P}_{auto}^{ex, c}(\omega)$, because energy does not change during tunneling. 

The connection between the auto-correlation noise and energy becomes even more transparent in the final example, where the dwell time is the only characteristic time. 

The last, third example is injection from a quantum dot with the equidistant ladder of levels, which is suddenly raised by one level spacing $ \Delta$ at $t=0$.\cite{Feve:2007jx,Filippone:2020fu} 
The Fermi level is exactly between the two successive levels. 
The probability of tunneling between the dot and the Fermi sea is small. 
The wave function, $\Psi^{ \Delta}(t) = e^{ - \frac{ i }{ \hbar  } \mu  t  }\psi^{ \Delta}(t)$, has an envelope \cite{Haack:2013ch,Moskalets:2013dl}

\begin{eqnarray}
\psi^{ \Delta}(t) &=& 
\theta(t) 
\frac{e^{- i \omega_{0} t } }{ \sqrt{ \tau_{D} } } 
e^{ -\frac{t }{2 \tau_{D} }  }  
.
\label{06}
\end{eqnarray}
\noindent \\
Here $ \theta(t)$ is the Heaviside step function and $ \hbar \omega_{ 0} = \Delta/2 $.  
Notice, in this case, the wave packet width is determined by the dwell time,  while the energy of an injected electron is $ \hbar \omega_{0}$, which is unrelated to $ \tau_{D}$.

For not too large frequencies, $ \omega \ll \Delta/ \hbar$, the straightforward calculations lead to (see Appendix~\ref{ap01} for details)

\begin{eqnarray}
{\cal P}_{auto}^{ex, \Delta}(\omega) &=& 
R \frac{ e^{2} }{ {\cal T} _{0} }  ,
\label{07} \\
{\cal P}_{cross}^{ \Delta}(\omega) &=& - 
R  \frac{ e^{2} }{ {\cal T} _{0} }  
\frac{ 1 }{ 1 + \left( \omega  \tau_{D} \right)^{2} }  .
\nonumber 
\end{eqnarray}
\noindent \\
Since the particles are injected far above the Fermi sea, $ \Delta/2 \gg \hbar \omega $, all of them contribute to noise. 
As a result, there is no energy related suppression. 
Therefore, the auto-correlation noise is independent on frequency. 
On the other hand, the density profile has a finite width, $ \tau_{D}$. 
Therefore,  the cross-correlation noise gets suppressed at $ \omega \geq \tau_{D}^{-1}$. 

One more important conclusion can be drawn from Eq.~(\ref{01}). 
If a single-particle envelope wave function, $ \psi$, is real-valued, for example, as in the case of a Majorana fermion,\cite{Beenakker:2011tp,Aguado:2017hc} its contribution to the auto-correlation noise is identically zero in the weak backscattering regime. 
I emphasize that this conclusion applies to traveling single particles in the Fermi sea, and not to localized states. 

It is also equal to zero if there is no wave splitter at all ($R=1$ in the present notation).  \cite{Hassler:2020te}

It worth to be mentioned, the charge conservation implies ${\cal P}_{auto}^{ex}\left(  0 \right) + {\cal P}_{cross}^{}\left(  0 \right) = 0$, see Appendix~\ref{nc}. \cite{Blanter:2000wi} 
This fact imposes some indirect constraint on the wave function of a single-electron wave packet that can be injected into a one-dimensional Fermi sea. 
In particular, no a single particle with a real (scalar) wave function can be injected in a clean manner, that is, without accompanying electron-hole pairs. 
Indeed, as the equation (\ref{01}) predicts, the cross-correlation noise at zero frequency is not zero, ${\cal P}_{cross}^{}\left(  0 \right) = - R e^{2} / {\cal T} _{0} \ne 0$.  
While in the case of a real-valued wave function, the excess  auto-correlation noise vanishes for any frequency is zero, ${\cal P}_{auto}^{ex}\left(  \omega \right) = 0$. 
To resolve seeming violation of the charge conservation, we must assume that if such a particle is injected, then the additional excitations are unavoidable created. 
The example is a half-leviton~\cite{{Moskalets:2016kc}}, a particle with a real wave function whose creation is accompanied by the creation of an electron-hole cloud. 

The rest of a paper is structured as follows. 
In Sec.~\ref{gen} the connection between the electrical current correlation functions, the auto- and cross-correlation noise power and the first-order correlation function of the periodic stream of electrons injected into a chiral conductor is established within the framework of the Floquet scattering matrix approach. 
Such a general connection allows for a detailed analysis of the similarities and differences between the auto- and cross-correlation noise, which is illustrated in Sec.~\ref{ex} using some examples. 
The conclusion is given in Sec.~\ref{concl}. 
Some technical details of calculations are presented in Appendices~\ref{ap01} - \ref{ap04} 

\section{Electrical noise and electron correlation function}
\label{gen}

To drive a current through a conductor, some external source is needed. 
The role of the source can be played, for example, by a constant or time-dependent voltage applied across a conductor, a time-dependent gate voltage, which changes the position of the quantum levels of electrons in the conductor or in its part, etc. 
If the characteristics of the source are known, the current can be calculated. 
In the quantum coherent regime, when the current is carried by individual electrons, the characteristics of carriers, for example, their wave function, are also can be calculated using the characteristics of the source. 
The measurements of electrical current and its fluctuations were already used to acquire information on quantum state of carriers.\cite{Jullien:2014ii,Bisognin:2020dk,Fletcher:2019jc}
Therefore, it is desirable to have a direct relation between the electrical and electron characteristics without explicit recursion to the characteristics of the  source. 
Some efforts in this direction have already been made.\cite{Moskalets:2015ub,Moskalets:2016fm,Moskalets:2020fm} 
Below the fluctuations of an electrical current are expressed in terms of the wave functions, more precisely, in terms of the excess first-order correlation function \cite{Grenier:2011js,Grenier:2011dv,Haack:2013ch,Moskalets:2013dl}  of electrons responsible for those fluctuations.   

To be specific, here I am interested in a quantum-coherent conductor connected via one-channel (chiral) leads \cite{Buttiker:2009bg} to several electron reservoirs in equilibrium.  
Some (or all) incoming leads are fed by external sources working periodically with period $ {\cal T} _{0}$. 

\subsection{Frequency-dependent noise}

The correlation function of currents, $I_{ \alpha}, I_{ \alpha ^{\prime}}$, flowing in leads $ \alpha$ and $\alpha ^{\prime}$ of a multi-terminal conductor is defined as follows,\cite{Blanter:2000wi} 

\begin{eqnarray}
{\cal P}_{ \alpha \alpha ^{\prime} }\left(  \omega \right) &=& \int\limits_{ -{\cal T} _{0}/2 }^{ {\cal T} _{0}/2 } \frac{dt }{ {\cal T} _{0} } 
\int\limits_{- \infty}^{ \infty } d \tau 
e^{i \omega \tau }
\Big\{ - I_{ \alpha}(t+ \tau) I_{ \alpha ^{\prime}}(t) 
\label{n01} \\
&& +
\frac{ 1 }{ 2 }
\left \langle  \hat I_{ \alpha}(t+ \tau)  \hat I_{ \alpha ^{\prime}}(t) +   \hat I_{ \alpha ^{\prime}}(t)  \hat I_{ \alpha}(t+ \tau) \right \rangle 
\Big\}
,
\nonumber 
\end{eqnarray}
\noindent \\
where $  \hat I_{ \alpha}$ and $ I_{ \alpha} =  \left \langle \hat I_{ \alpha} \right \rangle $ are an operator in second quantization and a corresponding measurable for a current in the lead $ \alpha$; the angular brackets $ \left \langle \dots \right \rangle$ denote the quantum statistical average; for a periodic drive, $ {\cal T} _{0}$ is a period, for a non-periodic drive $ {\cal T} _{0} \to \infty$. 
Note the difference in the factor of 2 compared to the definition used in Ref.~\onlinecite{Blanter:2000wi}

The current operator $\hat I_{ \alpha}$ is expressed in terms of creation and annihilation operators $\hat a_{ \alpha}^{ \dag}(E), \hat a_{ \alpha}(E)$ of electrons with energy $E$ incoming from the reservoir $ \alpha$ and  operators $\hat b_{ \alpha}^{ \dag}(E), \hat b_{ \alpha}(E)$ of electrons with energy $E$ scattered into the reservoir $ \alpha$.\cite{Buttiker:1990tn} 
In the wide band limit, that is, when the relevant energy scales, such as a  voltage applied, a temperature, the energy quantum $ \hbar \Omega$ with $ \Omega = 2 \pi/ {\cal T} _{0}$, etc., are all small compared to the Fermi energy $ \mu_{ \alpha}$, the current operator reads,\cite{Buttiker:1992vr} 

\begin{eqnarray}
\hat I_{ \alpha}(t) &=& \frac{ e }{ h  } \iint _{}^{ } dE dE ^{\prime}
e^{i \frac{ E - E ^{\prime} }{ \hbar } t }
\label{n02} \\
&& \times
\left\{ \hat b_{ \alpha}^{ \dag}(E) \hat b_{ \alpha}^{ }(E ^{\prime}) - \hat a_{ \alpha}^{ \dag}(E) \hat a_{ \alpha}^{ }(E ^{\prime}) \right\}.  
\nonumber 
\end{eqnarray}
\noindent \\
In the case of a periodically driven conductor, the operators $ \hat b_{ \alpha}$ are related to various operators $ \hat a_{ \alpha}$ via the elements of the unitary Floquet scattering matrix $S_{F}$,\cite{Moskalets:2002hu} 

\begin{eqnarray}
\hat b_{ \alpha}\left(  E \right) &=& 
\sum\limits_{ \beta}^{} 
\sum\limits_{n = - \infty}^{ \infty }
S_{F, \alpha \beta} \left(  E, E_{n} \right) \hat a_{ \beta}\left(  E_{n} \right) ,
\label{n03}
\end{eqnarray}
\noindent \\
where I introduced a short notation $E_{n} = E + n \hbar \Omega$. 
Charge conservation requires the scattering matrix to be unitary, which means,

\begin{eqnarray}
\sum_{ \gamma}^{}
\sum\limits_{n=-\infty}^{\infty}
S_{F, \gamma \alpha }^{*}\left(  E_{n}, E_{m} \right)
S_{F, \gamma \beta}^{}\left(  E_{n}, E_{} \right) = \delta_{ \alpha \beta} \delta_{m,0}, 
\nonumber \\ 
\label{n04} \\
\sum_{ \gamma}^{}
\sum\limits_{n=-\infty}^{\infty}
S_{F,  \alpha \gamma }^{*}\left(  E_{m}, E_{n} \right)
S_{F,  \beta \gamma}^{}\left(  E_{}, E_{n} \right) = \delta_{ \alpha \beta} \delta_{m,0}, 
\nonumber 
\end{eqnarray}
\noindent \\
where $ \delta_{n,0}$ is the Kronecker delta. 

Equation (\ref{n03}) allows to express the quantum-statistical average of the product of $b-$operators in terms of that of $a-$operators. 
Because the reservoirs are in equilibrium, the latter average is known. 
In the case of reservoirs of non-interacting electrons forming the Fermi sea, we have $\left \langle \hat a_{ \beta}^{ \dag}(E) \hat a_{ \beta}^{ }(E ^{\prime})  \right \rangle = f_{ \beta}(E) \delta\left(  E - E ^{\prime} \right)$, where $f_{ \beta}(E)$ is the Fermi distribution function with temperature $ \theta_{ \beta}$ and chemical potential $ \mu_{ \beta}$, and $  \delta\left(  E - E ^{\prime} \right)$ is the Dirac delta.

\subsubsection{$2 \times 2$ circuit}

Our aim is to compare auto- and cross-correlation noise. 
The minimal circuit that allows cross-correlation noise is an electronic wave splitter, a  quantum point contact (QPC) with two incoming, $ \beta = 1,2$, and two outgoing, $ \alpha = 3, 4$, channels, see Fig.~\ref{fig1}. 

Below I am interested in current fluctuations in out-going channels, that is, $ \alpha, \alpha ^{\prime} = 3,4$ in Eq.~(\ref{n01}). 
For this case, the general equation for noise within the Floquet scattering matrix approach \cite{Moskalets:2007dl,Moskalets:2011jx}  gives us,

\begin{subequations}
\label{n06-1} 
\begin{eqnarray}
{\cal P}_{33}^{}\left(   \omega \right)= 
\frac{e^2}{h}\int\limits_{}^{}dE
\Big\{
F_{33}(E,E+\hbar\omega)
+ 
\sum\limits_{n,m,q=- \infty}^{ \infty}
\nonumber \\ 
\sum\limits_{\delta=1}^{2}
\sum\limits_{\gamma=1}^{2}
F_{ \gamma \delta}(E_q + \hbar\omega, E_{}) 
S_{F,3\delta}^{*}(E_{n},E_{}) 
S_{F,3\delta}(E_{m}, E_{})
\nonumber \\
\times
S_{F,3\gamma}^{*}(E_{m}+\hbar\omega,E_{q}+\hbar\omega)
S_{F,3\gamma}(E_{n}+\hbar\omega, E_q+\hbar\omega) 
\Big\}
,
\nonumber \\
\label{n06-1a} 
\end{eqnarray}
\ \\ \noindent
and 

\begin{eqnarray}
{\cal P}_{34}(\omega) = 
\frac{e^2}{h}\int\limits_{}^{}dE
\sum\limits_{n,m,q=- \infty}^{ \infty}
\nonumber \\
\sum\limits_{\delta=1}^{2}
\sum\limits_{\gamma=1}^{2}
F_{ \gamma \delta}(E_q + \hbar\omega, E_{}) 
S_{F,3\delta}^{*}(E_{n},E_{}) 
S_{F,4\delta}(E_{m}, E_{})
\nonumber \\
\times
S_{F,4\gamma}^{*}(E_{m}+\hbar\omega,E_{q}+\hbar\omega)
S_{F,3\gamma}(E_{n}+\hbar\omega, E_q+\hbar\omega) ,
\nonumber \\
\label{n06-1b}
\end{eqnarray}
\ \\ \noindent
\end{subequations}
with 

\begin{eqnarray}
F_{ \gamma \delta}(E_1, E_{}) = \frac{ f_{ \gamma}\left( E_1  \right) + f_{ \delta} \left(  E \right)  }{ 2 } - f_{ \gamma}\left( E_1 \right) f_{ \delta} \left(  E \right) .
\label{n07}
\end{eqnarray}
\noindent \\
Let us also introduce excess noise, that is, an increase in noise due to the source, which is defined as the following difference  

\begin{eqnarray}
{\cal P}_{ \alpha \alpha ^{\prime} }^{ex}\left(   \omega \right) = {\cal P}_{\alpha \alpha ^{\prime}}^{}\left(  \omega \right) - {\cal P}_{\alpha \alpha ^{\prime}}^{off}\left(  \omega \right) ,
\label{exn}
\end{eqnarray}
\noindent \\
where the upper index ${off}$ indicates that the source is off. 

I suppose a unitary $2 \times 2$ scattering matrix of the QPC to be energy-independent, 

\begin{eqnarray}
S^{QPC} =
\left(
\begin{array}{ccc}
 \sqrt{R} &  i \sqrt{T}    \\
 i \sqrt{T} &  \sqrt{R}  
\end{array}
\right) ,
\label{04n-1}
\end{eqnarray}
\noindent \\
a real number $0 \leq R \leq 1$ is the reflection probability, the transmission probability $T = 1 - R$. 
We need an energy independent $S^{QPC}$ to use the noise to get information on injected wave packets only. 
If the properties of the electronic circuit that connects the incoming and outgoing channels depend on energy, the outgoing signal also carries nontrivial information about this circuit. \cite{Burset:2019ha}

In addition, for the sake of simplicity, I suppose that the periodic source is present only in the incoming channel $ \beta = 1$. 
It is characterized by the Floquet scattering amplitude, which is a matrix in an energy space with elements $S_{F}\left(  E, E_{n} \right)$.  
The results presented below can be directly generalized to the case when  another source is added in the second incoming channel, see Appendix~\ref{ap02-4}. 

For the circuit with single source and single QPC, the elements of the total Floquet scattering matrix are represented as follows, 

\begin{eqnarray}
S_{F, 3 1} \left(  E, E_{n} \right) &=& \sqrt{R} S_{F}\left(  E, E_{n} \right) ,
\nonumber \\
S_{F, 4 1} \left(  E, E_{n} \right) &=& i\sqrt{T} S_{F}\left(  E, E_{n} \right) ,
\nonumber \\
\label{n05} \\
S_{F, 3 2} \left(  E, E_{n} \right) &=& i\sqrt{T} \delta_{n,0},
\nonumber \\
S_{F, 4 2} \left(  E, E_{n} \right) &=& \sqrt{R} \delta_{n,0}.
\nonumber 
\end{eqnarray}
\noindent \\
All other elements are zero.

\subsection{First-order correlation function}

To characterize a quantum state injected by the source into a ballistic one-dimensional electronic waveguide, I use the first-order correlation function, $ {\cal G} ^{(1)}_{}$.
This function is defined as a quantum statistical average of the product of two field operators for electrons calculated in the electronic wave-guide $ \beta$ just after the source,
$
{\cal G} ^{(1)}_{ \beta}\left(  t_{1}; t_{2} \right) = \left \langle \hat \Psi ^{ \dag}_{ \beta}(t_{1}) \hat \Psi_{ \beta}(t_{2} )  \right \rangle 
$. \cite{Grenier:2011js} 
Strictly speaking, this object is a $2 \times 2$ matrix in the spin space. 
However, here I consider the spin-polarized case and suppress the spin index. 

In the case under consideration, the source is placed in lead $ \beta = 1$ and it is characterized by the Floquet scattering amplitude, $S_{F}\left(  E_{n}, E \right)$. 
The corresponding correlation function is calculated as follows, \cite{Moskalets:2015ub}

\begin{eqnarray}
v_{ \mu}{\cal G}^{(1)}_{1 }(t_{}+ \tau;t_{}) = \frac{ 1 }{ h }
\int dE  f_{1 }\left( E \right) 
e^{ \frac{ i }{ \hbar  } E  \tau } 
\sum_{n,m=-\infty}^{\infty} 
e ^{i \Omega  n \tau}
\nonumber \\
\times
e ^{i \Omega \left( n - m  \right) t}
S_{F}^{*}\left(E_{n},E  \right)
S_{F}\left(E_{m},E  \right)
. 
\quad
\label{cf01} 
\end{eqnarray}
\ \\ \noindent
Here $v_{ \mu}$ is a velocity of electrons at the Fermi level, which is originated from the density of states being energy independent in the wide band limit used here.

When the source is switched off, $S_{F}\left(  E_{n}, E \right) = \delta_{ n,0}$, the above equation is reduced to the correlation function of the Fermi sea in equilibrium,  $\left[ {\cal G} ^{(1)}_{0, \beta }( \tau) \equiv {\cal G} ^{(1)}_{0, \beta }(t_{}+ \tau;t_{}) \right]$

\begin{eqnarray}
v_{ \mu} {\cal G} ^{(1)}_{0, \beta }( \tau ) &=& \frac{ 1 }{ h }
\int  dE  f_{ \beta }\left( E \right) 
e^{ i \frac{ E }{ \hbar  }   \tau  } 
\nonumber \\
&=&
\frac{e ^{ i \tau \frac{ \mu_{ \beta}  }{ \hbar } }  }{ 2 \pi i  } 
\frac{ 1/  \tau_{ \theta_{ \beta} }}{ \sinh\left(  \tau /  \tau_{ \theta_{ \beta}  }    \right) } .
\label{cf01-1}  
\end{eqnarray}
\ \\ \noindent
Here $ \tau_{ \theta_{ \beta}} = \hbar /( \pi k_{B} \theta_{ \beta})$ is  the thermal coherence time for the reservoir, where the lead $ \beta$ is attached to.   

The difference of correlation functions with the source being on and off is the excess correlation function, which characterizes what is injected by the source into an electron waveguide $ \beta = 1$, \cite{Grenier:2011dv}

\begin{eqnarray}
{G}^{(1)}_{1 }\left(  t_{1}; t_{2} \right) &=& {\cal G}^{(1)}_{1 }\left(  t_{1}; t_{2} \right) - {\cal G}^{(1)}_{0,1 }\left(  t_{1} - t_{2} \right). 
\label{cf01-2}
\end{eqnarray}
\noindent \\
In the case when the source injects a single electron with wave function $ \Psi(t)$ per period, the excess correlation function during that period takes on a very simple form, ${G}^{(1)}_{1 }\left(  t_{1}; t_{2} \right) = \Psi^{*}\left(  t_{1} \right) \Psi\left(  t_{2} \right)$. 
In the case of a multi-particle injection  ${G}^{(1)}_{1 }\left(  t_{1}; t_{2} \right) = \sum_{j}^{} \Psi_{j}^{*}\left(  t_{1} \right) \Psi_{j}\left(  t_{2} \right)$. 
\cite{{Grenier:2013gg}}

\subsection{Noise power in terms of ${\cal G} ^{(1)}_{}$}
\label{2c}

In the case of non-interacting electrons, the correlation function $ {\cal G} ^{(1)}_{}$ contains complete information about the system of electrons.  
In particular, all measurables can be expressed in terms of correlation function, see, e.g., Ref.~\onlinecite{Moskalets:2015ub} for some examples. 
Such expressions are notably useful when transport is due to only a few electrons per period. 

Here I express the excess noise in terms of $G ^{(1)}_{}$ in the case when all  incoming channels have the same temperature $ \theta_{ \beta} = \theta$ and Fermi energy $ \mu_{ \beta} = \mu$. 
Therefore, the equilibrium electronic correlation functions are the same, ${\cal G} ^{(1)}_{0, \beta } = {\cal G} ^{(1)}_{0}$. 
For more general case and for details of calculations, see Appendix \ref{ap02}. 

First, let us substitute Eqs.~(\ref{n05}) and (\ref{n07}) into Eqs.~(\ref{n06-1}) and calculate the excess noise, Eq.~(\ref{exn}). 
Then let us separate linear and bilinear in Fermi functions terms. 
The former ones are nullified, while the lather ones are expressed in terms of the correlation functions presented in Eqs.~(\ref{cf01}), (\ref{cf01-1}), and (\ref{cf01-2}).    
As a result, I find for the auto-correlation noise, [see Eq.~(\ref{ap0211-1})]

\begin{subequations}
\label{npg01} 
\begin{eqnarray}
{\cal P}_{ 33}^{ex}\left(  \omega \right)  &=& 
e^{2}  v_{ \mu}^{2} 
 \int\limits _{ - {\cal T} _{0}/2 }^{ {\cal T} _{0}/2 } \frac{dt }{ {\cal T} _{0} }  
\int\limits _{- \infty}^{ \infty } d \tau e^{i \omega \tau}  
\Bigg\{
\label{npg01a} \\
&&
- R^{2} \left | { G}_{1}^{(1)}\left(  t + \tau; t \right) \right |^{2} 
\nonumber \\
&&
- 2 R \,  {\rm Re} { G}_{1}^{(1)}\left(  t + \tau; t \right)  {\cal G}_{0}^{(1)*}\left( \tau \right) 
\Bigg\},
\nonumber 
\end{eqnarray}
\noindent \\ 
and for the cross-correlation noise, [see Eq.~(\ref{ap0215})]

\begin{eqnarray}
{\cal P}_{34}^{}(\omega)  &=& 
- 
R T 
e^{2} v_{ \mu}^{2}  
 \int\limits _{ - {\cal T} _{0}/2 }^{ {\cal T} _{0}/2 } \frac{dt }{ {\cal T} _{0} }  
\int\limits _{- \infty}^{ \infty } d \tau e^{i \omega \tau}  
\nonumber \\
&& \times
\left |  { G}_{1}^{(1)}\left(  t + \tau; t \right)   \right |^{2} .
\label{npg01b}
\end{eqnarray}
\end{subequations}
\noindent \\
Notice that when all reservoirs are in the same conditions, cross-correlation  noise disappears when the source is turned off. 
This is why the superscript ${ex}$ is omitted.

The important difference between auto- and cross-correlation noise, Eqs.~(\ref{npg01a}) and (\ref{npg01b}), is that the latter one is determined solely by what is injected by the source, while the former one in addition depends explicitly on the properties of the Fermi sea. 

The part of the noise that is determined by $\left | G ^{(1)}_{1} \right |^{2}$ depends on the possible quantum exchange~\cite{Blanter:2000wi} between the  injected electrons. 
At zero temperature and when electrons are injected one at a time without overlapping, this part of the noise is reduced to the product of currents in Eq.~(\ref{n01}).   
In the wide band approximation used here, the electric current is proportional to the density profile of the wave packet, hence Eq.~(\ref{01}), the second line. 

In contrast, the part of the noise that is determined by the product of $G ^{(1)}_{1}$ and ${\cal G} ^{(1)}_{0}$  takes into account the quantum exchange of an injected electron and electrons of the Fermi sea.  
Such an exchange does not contribute to cross-correlation noise, unless the two incoming Fermi seas are different, see Eq.~(\ref{ap0214}).  

The formal difference between the auto- and cross-correlation noise becomes especially pronounced  in the weak back-scattering regime, $R \ll 1$, when we can discard the terms \( \sim R^{2} \) in Eqs.~(\ref{npg01}) and get the following,

\begin{eqnarray}
{\cal P}_{ 33}^{ex}\left(  \omega \right)  &=& 
- R e^{2}  v_{ \mu}^{2} 
 \int\limits _{ - {\cal T} _{0}/2 }^{ {\cal T} _{0}/2 } \frac{dt }{ {\cal T} _{0} }  
\int\limits _{- \infty}^{ \infty } d \tau e^{i \omega \tau}  
\nonumber \\
&&
2  {\rm Re} { G}_{1}^{(1)}\left(  t + \tau; t \right)  {\cal G}_{0}^{(1)*}\left(  \tau\right) ,
\nonumber \\
\label{npg02} \\
{\cal P}_{34}^{}(\omega)  &=& - R e^{2} v_{ \mu}^{2}  
 \int\limits _{ - {\cal T} _{0}/2 }^{ {\cal T} _{0}/2 } \frac{dt }{ {\cal T} _{0} }  
\int\limits _{- \infty}^{ \infty } d \tau e^{i \omega \tau}  
\left |  { G}_{1}^{(1)}\left(  t + \tau; t \right)   \right |^{2} .
\nonumber 
\end{eqnarray}
\noindent \\
Using Eq.~(\ref{cf01-1}) for ${\cal G} ^{(1)}_{0}$ at zero temperature, $ \theta = 0 \Rightarrow \tau_{ \theta} \to \infty$,  and for a single-particle injection, $ G ^{(1)}_{}\left(  t + \tau; t \right)  = e^{ i \frac{ \mu }{ \hbar  } \tau  } \psi^{*}\left(  t + \tau \right) \psi(t)$, I arrive at Eq.~(\ref{01}) with ${\cal P}_{auto}^{ex} = {\cal P}_{33}^{ex}$ and ${\cal P}_{cross}^{} = {\cal P}_{34}^{}$, where the integration over $t$ is extended to infinity, because the duration of wave packet is much less than the period $ {\cal T} _{0}$.

\section{Examples}
\label{ex}

Here I will consider two examples: one when auto- and cross-correlation noise are perfectly anti-correlated at any frequency, and the other when they can be   different.

\subsection{Energy-independent source}

In the case when the properties of the source do not change on the scale of the energy of the injected particles, the corresponding Floquet scattering amplitude can be represented as a Fourier coefficient of a certain energy-independent scattering amplitude, $S_{F}\left(  E_{n}, E \right) = \int _{0 }^{ {\cal T} _{0} } \frac{ dt }{ {\cal T} _{0} } e^{2 \pi i n \frac{ t }{ {\cal T} _{0}  } } S(t)$. \cite{Moskalets:2002hu,Moskalets:2013dl}
In a one-dimensional case, unitarity implies that $S(t)$ is a pure phase factor, that is, $\left | S(t) \right |^{2} = 1$. 
For example, if a voltage $V(t)$ plays a role of a source, then it is $S(t) = \exp\left(  i \frac{ e }{ \hbar  } \int _{}^{ t } d t ^{\prime} V(t ^{\prime}) \right)$.  

In such a case, equation (\ref{cf01}) gives us, ${\cal G}^{(1)}_{1 }(t_{1};t_{2}) = S^{*}(t_{1}) S(t_{2}) {\cal G}^{(1)}_{0,1 }(t_{1}-t_{2})$.
Remind that ${\cal G}^{(1)}_{0,1 }$ describes the Fermi sea in equilibrium at temperature $ \theta$. 
Using this result in Eqs.~(\ref{npg01}) and taking into account that  

\begin{eqnarray}
{ G}^{(1)}_{1 }(t_{1};t_{2}) = \left\{ S^{*}(t_{1}) S(t_{2}) - 1 \right\} {\cal G}^{(1)}_{0,1 }(t_{1}-t_{2}), 
\label{ex01}
\end{eqnarray}
\noindent \\ 
I find that the excess auto- and cross-correlation noise are perfectly anti-correlated at any frequency, 

\begin{eqnarray}
{\cal P}_{ 33}^{ex}\left(  \omega \right) + {\cal P}_{ 34}^{}\left(  \omega \right) = 0 ,
\label{ex02}
\end{eqnarray}
\noindent \\
not only at zero frequency, $ \omega = 0$, as the charge conservation requires \cite{Blanter:2000wi}, see also Appendix~\ref{nc}. 
An example was shown in Eq.~(\ref{03}). 

For $R=1$, when  cross-correlation noise does not exist, that is, formally ${\cal P}_{ 34}^{} = 0$, the above equation tells us that whatever emitted by the source under consideration is silent on any frequency, the excess auto-correlation noise is zero, ${\cal P}_{ 33}^{ex}\left(  \omega \right) = 0$.   
In particular, any voltage applied to the ballistic channel produces no excess noise at any frequency. 
Note that for a generic source injecting electrons into a ballistic waveguide, a  similar conclusion can be drawn for noise only at zero frequency, see Eq.~(\ref{npg03}). 

Some general conclusions can be made regarding  the effect of temperature on noise. 
Indeed, the equations (\ref{ex01}) and (\ref{cf01-1}) allow us to relate the excess correlation function at zero (the extra subscript $0$) and non-zero (the extra subscript $ \theta$) temperatures, \cite{}

\begin{eqnarray}
{ G}^{(1)}_{1, \theta }\left(  t+ \tau; t \right) = 
\frac{ \tau/  \tau_{ \theta_{ } }}{ \sinh\left(  \tau /  \tau_{ \theta_{ }  }    \right) } 
\, { G}^{(1)}_{1, 0 }\left(  t+ \tau; t \right) .
\label{ex03}
\end{eqnarray}
\noindent \\
Then I use the above equation in Eqs.~(\ref{npg01}), utilize the inverse Fourier transformation with respect to $ \omega$ and express the  noise at temperature $ \theta$, ${\cal P}_{ \theta}^{ex}$,  in terms of the noise at zero temperature, ${\cal P}_{0}^{ex}$, as follows, 

\begin{eqnarray}
{\cal P}_{ \theta}^{ex}\left(  \omega \right) \!\! &=& \!\! 
\int\limits _{- \infty}^{ \infty } 
\!\!
d \tau e^{i \omega \tau} 
\left(  \frac{ \tau/  \tau_{ \theta_{ } }}{ \sinh\left(  \tau /  \tau_{ \theta_{ }  }    \right) }  \right)^{2}
\!\!
\int\limits _{- \infty}^{ \infty } \frac{ d \omega ^{\prime} }{ 2 \pi  } 
e^{ - i \omega ^{\prime} \tau} 
{\cal P}_{0}^{ex}\left(  \omega ^{\prime} \right) 
.
\nonumber \\
\label{ex04}
\end{eqnarray}
\noindent \\
Here I introduce ${\cal P}^{ex} \equiv {\cal P}_{33}^{ex} = - {\cal P}_{34}^{} $ according to Eq.~(\ref{ex02}).

\subsection{Injection from a quantum level raising at a constant rapidity}

Now let us consider a single-electron injection from a source, whose properties do depend on energy. 
The corresponding scattering amplitude and the wave function of the injected electron were discussed in Ref.~\onlinecite{Keeling:2008ft} at zero temperature and in Ref.~\onlinecite{Moskalets:2017fh} at nonzero temperatures. 
In this case, the auto- and cross-correlation noises don't stick together unless at zero frequency.

At zero temperature, the excess correlation function is $ v_{ \mu} G ^{(1)}_{1}\left(  t+ \tau; t \right) = e^{i \frac{ \mu }{ \hbar  } \tau  } \psi^{(c)*}\left(  t+ \tau \right) \psi^{(c)}\left(  t \right)$, where  $ \psi^{(c)}$ is shown in Eq.~(\ref{04}). 
Using this equation in Eqs.~(\ref{npg01}) and assuming that the width of the wave packet is small compared to the period, $ \Gamma _{\tau} \ll {\cal T} _{0}$, I find, (see Appendix \ref{ap03} for details)

\begin{eqnarray}
{\cal P}_{33}^{ex}(\omega) &=& 
R
\frac{ e^{2} }{ {\cal T} _{0} }   
\frac{ T + \left( \omega  \tau_{D} \right)^{2}  }{ 1 + \left( \omega  \tau_{D} \right)^{2} } e^{-  \left | \omega \right |  2 \Gamma _{\tau}} ,
\nonumber \\
{\cal P}_{34}^{}(\omega) &=& - 
R \frac{ e^{2} }{ {\cal T} _{0} } 
\frac{ T }{ 1 + \left( \omega  \tau_{D} \right)^{2} } 
e^{- \left | \omega \right |  2 \Gamma _{\tau}} 
.
\label{c01} 
\end{eqnarray}
\ \\ \noindent
At $R \ll 1$ ($T \approx 1$) we reproduce Eq.~(\ref{05}).

\begin{figure}[t]
\includegraphics[width=10.5 cm]{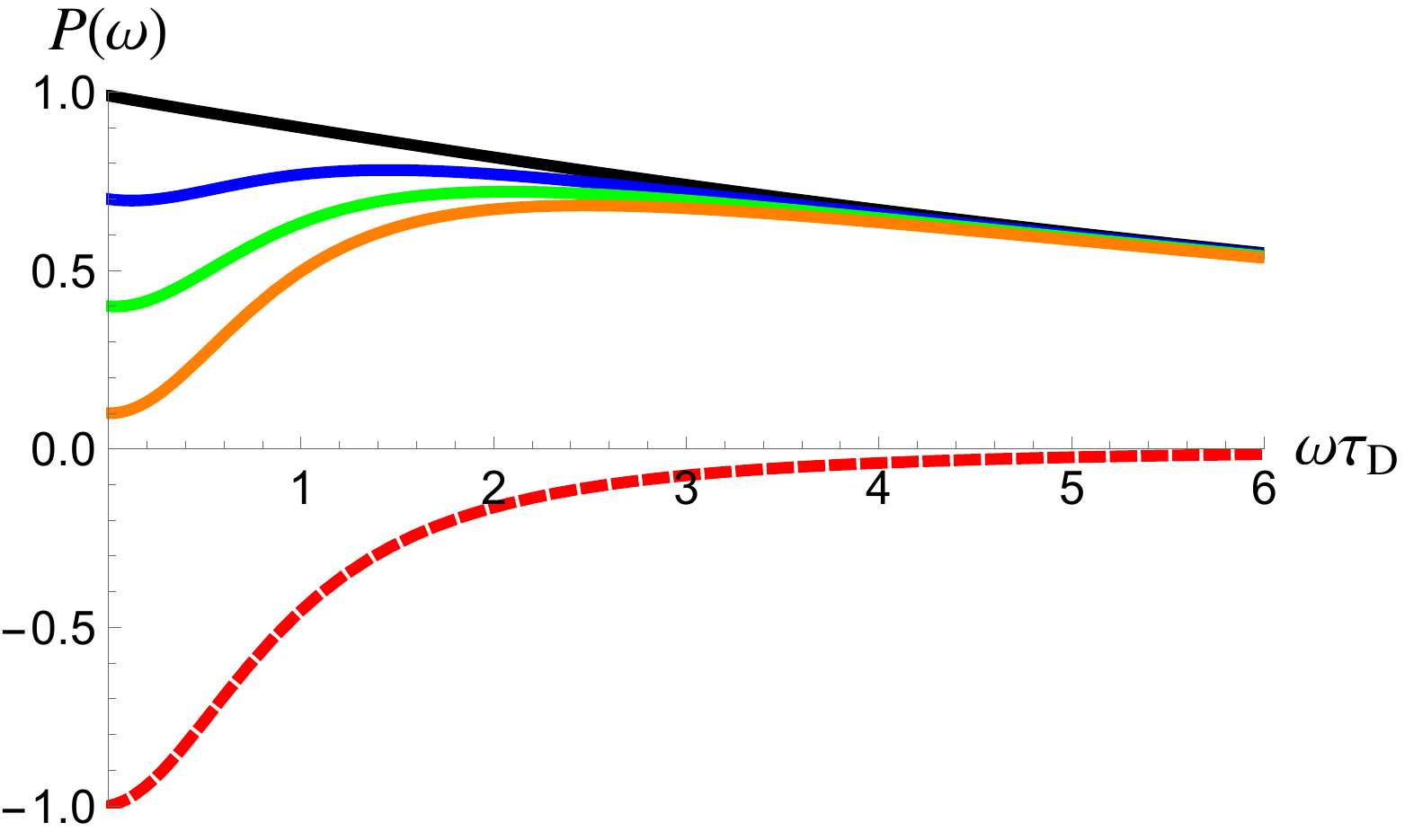}
\caption{
The excess auto- (solid lines) and cross-correlation (a red dashed line) noise are  shown as a function of the frequency $ \omega$ at zero temperature, see Eq.~(\ref{c01}). 
The cross-correlation noise, ${\it P}_{}^{} \equiv {\cal P}_{34}^{}$, is given in units of 
$ RT e^{2}/ {\cal T} _{0}$. 
The excess auto-correlation noise, ${\it P}_{}^{} \equiv {\cal P}_{33}^{ex}$, is given in units of $R e^{2}/ {\cal T} _{0}$ for $T=0.999$ (a black line), $T=0.7$ (a blue line), $T=0.4$ (a green line), and $T=0.1$ (an orange line). 
The parameter $2 \Gamma _{\tau}/ \tau_{D} = 0.1$. 
}
\label{fig2}
\end{figure}

The above equations are illustrated in Fig.~\ref{fig2} in the case of $2 \Gamma _{\tau} \ll  \tau_{D}$, when the difference between them is most pronounced. 
As I already discussed in Introduction after Eqs.~(\ref{05}), the excess auto-correlation  noise and the cross-correlation noise demonstrate significantly  different dependences on frequency. 
The  cross-correlation noise (its absolute value) decreases monotonically with frequency, see Figure~\ref{fig2}, a red dashed line. 
In contrast, the excess auto-correlation noise is non-monotonically dependent on frequency, which is a manifestation of the existence of two contributions. 
The first contribution, which is responsible for the quadratic increase at low  frequencies, is similar to the cross-correlation noise, compare the first term in  Eq.~(\ref{npg01a}) and Eq.~(\ref{npg01b}). 
While the second contribution is different. 
This contribution dominates in the limit of $T \to 1$ and at high frequencies, see a black solid line in Figure~\ref{fig2}. 

At non-zero temperature, auto- and cross-correlation noise are modified by the same factor, (see Appendix \ref{ap04} for details),

\begin{eqnarray}
\eta_{ }\left(   \omega, \theta \right) = 
e^{  \left | \omega \right | 2 \Gamma _{\tau}}  
\int\limits_{- \infty}^{ \infty } \frac{ d x }{  \pi  }  
\frac{ 
e^{i x \omega 2 \Gamma _{\tau} }
 }{x^{2} + 1 } \left(  \frac{ x 2 \Gamma _{\tau} /  \tau_{ \theta_{ } }}{ \sinh\left(  x 2 \Gamma _{\tau} /  \tau_{ \theta_{ }  }    \right) } \right)^{2}
.
\label{c02}
\end{eqnarray}
\noindent \\
Interestingly, the above equation is independent of the dwell time $ \tau_{D} = \hbar /\left(  2 \Gamma _{\tau} c \right)$, where $c$ is a rapidity, see Eq.~(\ref{04}). 
Therefore, the same factor $ \eta_{ }\left(  \omega, \theta \right)$ applies for the case of $ \tau_{D} = 0$, which is the case for the source of levitons of half-width $ \Gamma _{\tau}$.  
However this analogy is not complete. 

Namely, for levitons the high-temperature noise can be expressed directly in terms of the noise at zero temperature, see Eq.~(\ref{ex04}). 
On the contrary, for electrons emitted from the quantum level, this is generally not the case due to factors depending on $ \omega \tau_{D}$. 
Yet, in the weak backscattering regime, $R \ll 1$, the auto-correlation noise obeys Eq.~(\ref{ex04}), because it does not depend on $ \tau_{D}$, see Eq.~(\ref{05}).

\begin{figure}[t]
\centering
\resizebox{0.99\columnwidth}{!}{\includegraphics{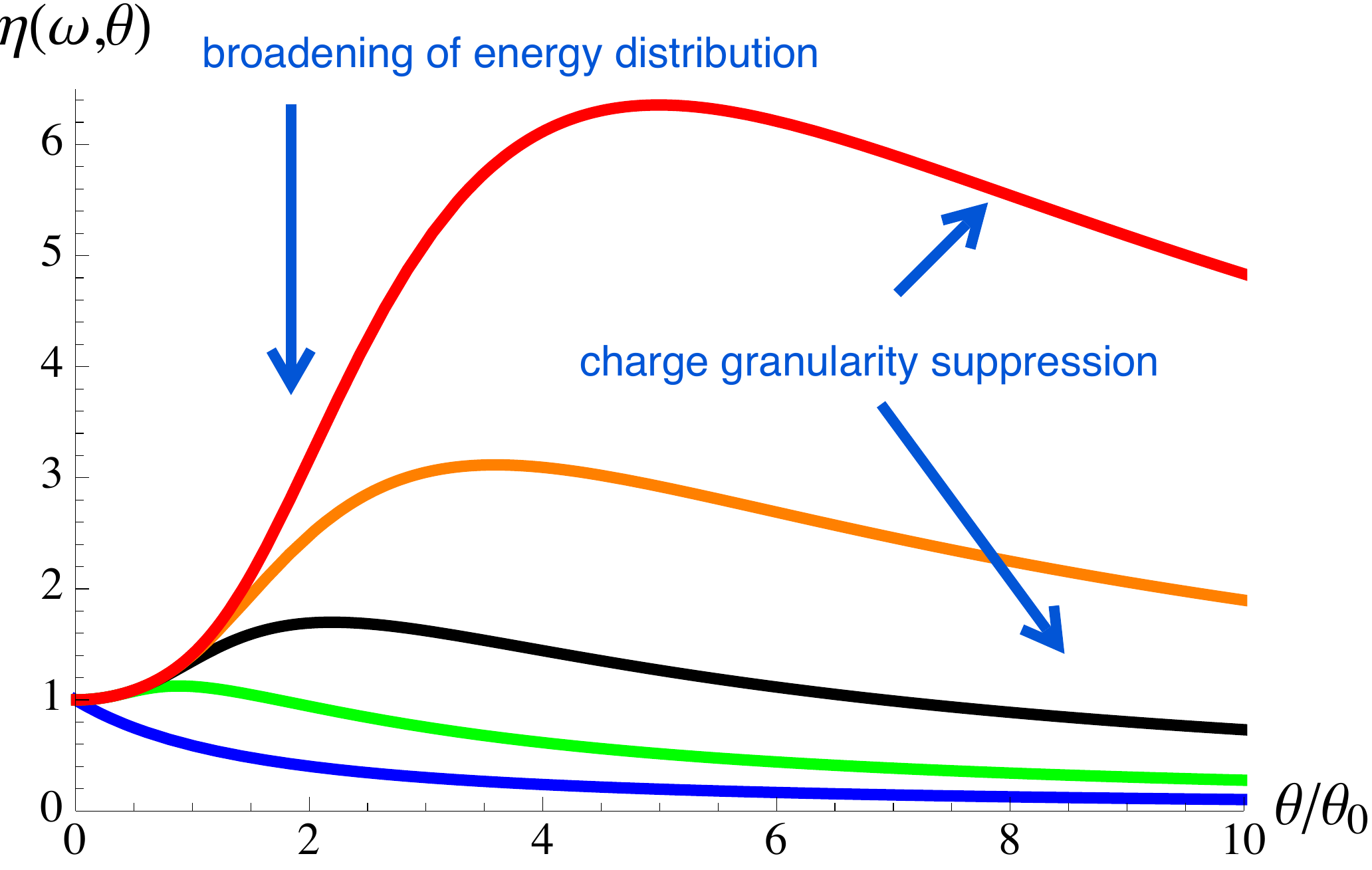} }
\caption{
The factor  $ \eta\left(  \omega, \theta \right)$, Eq.~(\ref{c02}), is shown as a function of temperature $ \theta$ for $ \omega = n/(2 \Gamma _{\tau})$ with $n=0\,({\rm blue})$, $n=1\,({\rm green})$, $n=2\,({\rm black})$, $n=3\,({\rm orange})$, and $n=4\,({\rm red})$.  
The parameter $ \theta_{0}$ is the temperature when the thermal coherence length is equal to the width of the wave packet, $2 \Gamma _{\tau} = \hbar /( \pi k_{B} \theta_{ 0})$. 
}
\label{fig3}
\end{figure}

The temperature-dependent factor $ \eta\left(  \omega, \theta \right)$, Eq.~(\ref{c02}), is shown in Fig.~\ref{fig3} for several frequencies $ \omega$. 
Remarkably, the maximum occurs at $ \omega \tau_{ \theta} \sim 1$, which is independent of properties of the source.  
The non-monotonic temperature behavior at non-zero frequencies is due to  two counter acting effects, both due to the fact that the quantum state of electrons injected at non-zero temperatures is a mixed quantum state.\cite{Moskalets:2015ub} 

The first effect, which leads to noise suppression, comes from the fact that each component of a mixed state is scattered independently at the wave splitter. 
Such an independent scattering manifests itself as second order coherence. \cite{Moskalets:2018ia}
As a result, the effect of charge quantization becomes less pronounced, and shot noise decreases with increasing temperature. 
At $ \theta \gg \theta_{0}$, where $k_{B} \theta_{0} = \hbar /( 2 \pi  \Gamma _{\tau})$, the shot noise decays as $ \theta_{0}/ \theta$. \cite{Moskalets:2017dy}  

Suppression of a zero-frequency shot noise with temperature has been reported in Refs.~\onlinecite{Bocquillon:2012if,Bocquillon:2013fp,Glattli:2016wl,Glattli:2016tr}. 
For the source of levitons, this effect was predicted in Ref.~\onlinecite{Dubois:2013fs}. 

The second effect, leading to an increase in noise, is associated with an  effective broadening of the energy distribution of injected particles caused by broadening of the probability density for the components of mixed state with increasing temperature, $p( \epsilon) =   -  \partial f_{1} ( \epsilon) /  \partial \epsilon $, see Eq.~(\ref{t01}). 
As a result, the injected particle is more likely to be able to emit energy $ \hbar \omega$ in order to contribute to noise at frequency $ \omega$. \cite{Parmentier:2012ed} 
This increase reaches saturation at $4 k_{B} \theta \sim \hbar \omega$, which leads to the maxima in Fig.~\ref{fig3}.

\section{Conclusion}
\label{concl}

The finite-frequency fluctuations of an electric current in multi-terminal conductors were analyzed at zero as well as non-zero temperatures. 
The focus was on the quantum coherent regime, when the  current is carried by non-overlapping single-particle wave packets periodically injected into a unidirectional, chiral wave guide. 

To highlight similarities and differences between auto- and cross-correlation noise, the fluctuations of an electric current were expressed in terms of the wave functions of injected electrons, bypassing the use of explicit source  characteristics. 
Two contributions to shot noise have been identified. 
The first, which depends on the possible quantum exchange between the injected electrons, determines the cross-correlation noise and part of the auto-correlation noise. 
In the case of single-particle injection and at zero temperature, this part is determined by the density profile of the injected wave packets. 
The second contribution, which depends on the quantum exchange of an injected electron and electrons of the Fermi sea, contributes only to the auto-correlation noise.  
This part is determined by the  coherence of the injected wave packets multiplied by the coherence of the Fermi sea.

At zero frequency, the charge conservation tightly links both contributions to shot noise. 
Such a connection allows us to make some general conclusions related to the properties of excitations that can be injected/created in the Fermi sea. 
In particular, no excitations with a real wave function can be created in the Fermi sea without accompanying electron-hole pairs, which follows from the fact that otherwise Eq.~(\ref{01}) would be incompatible with the conservation of charge \cite{Blanter:2000wi}, see Eq.~(\ref{ex02}) at $ \omega=0$. 

At non-zero frequencies, the two contributions in question are generally different. 
They can be addressed separately by measuring both auto- and cross-correlation noise in the weak backscattering regime, $R \ll 1$, when electrons are rarely scattered into the detector.  
On contrary, at $R=1$, the auto-correlation noise, the phase noise \cite{Mahe:2010cp} is determined by the combination of both contributions.  

For several experimentally available single-electron sources for which the wave function was calculated, I compared auto- and cross-correlation noise. 
For the family of so called energy-independent sources, the source of leviton \cite{Dubois:2013ul} is an example, the contribution related to the density profile and the contribution related to the coherence of the wave packet  turn out to be the same, see Eq.~(\ref{03}). 
So, the auto- and cross-correlation noises are the same (up to the minus sign) at any frequency not only at zero frequency as the charge conservation requires, see Eq.~(\ref{ex02}),. 
For another source, which relies on tunneling, \cite{Feve:2007jx} the two contributions are manifestly different, see Eqs.~(\ref{05}) and (\ref{07}) for different working regimes of the source.  

I analyzed the effect of temperature on shot noise in the case when electrons are injected on top of the Fermi. 
It turns out that temperature affects both contributions equally, see Eq.~(\ref{c02}) for the temperature-dependent factor. 
Since one of the contributions to shot noise depends on the quantum state of electrons in the Fermi sea, I conclude that temperature affects the quantum state of both the electrons in the Fermi sea and the injected electrons in the same way. 
Namely, a pure state at zero temperature becomes a mixed state at non-zero temperatures.\cite{Moskalets:2015ub}
At zero frequency, changing the quantum state from pure to mixed leads to  noise suppression, while at non-zero frequencies, the temperature dependence of noise is non-monotonic, see Fig.~\ref{fig3}.  
The temperature-dependent factor peaks when the thermal coherence time becomes of the order of the inverse of the frequency at which the noise is measured. 
Importantly, the position of this maximum does not depend on the properties of  the sources under consideration.

\acknowledgments
I acknowledge the warm hospitality and support of Tel Aviv University, and support from the Ministry of Education and Science of Ukraine (project No. 0119U002565). 
I am grateful to Pascal Degiovanni for stimulating discussions.

\appendix

\section{Auto-correlation noise from Eq.~(\ref{07})}
\label{ap01} 

\begin{figure}[t]
\centering
\resizebox{0.99\columnwidth}{!}{\includegraphics{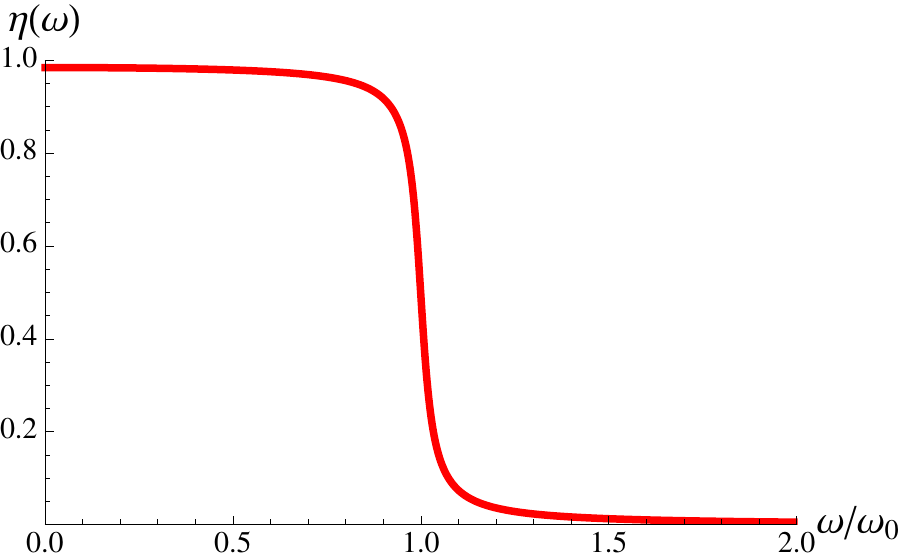} }
\caption{
A frequency-dependent suppression factor $ \eta( \omega)$, Eq.~(\ref{ap102}). 
The frequency $ \omega$ is given in units of $ \omega_{0} = \Delta/ (2 \hbar)$, see Eq.~(\ref{06}). 
The product $ \omega_{0} \tau_{D} = 20.$
}
\label{fig4}
\end{figure}

After substituting Eq.~(\ref{06}) into Eq.~(\ref{01}), the first line, I get, 

\begin{eqnarray}
{\cal P}_{auto}^{ex}\left(  \omega \right) &=& R
\frac{ e^{2} }{ {\cal T} _{0} }
\int\limits_{ 0}^{ \infty } \frac{ dt }{ \tau_{D}  } 
\int\limits_{ -t}^{ \infty } d \tau
e^{i \omega \tau }  
\frac{ \sin\left(\omega_{0}  \tau \right) }{ \pi \tau }
e^{ -\frac{t + \tau/2 }{\tau_{D} }  }  .
\nonumber \\
\label{ap101} 
\end{eqnarray}
\ \\ \noindent
To integrate with respect to $t$, I split the integration area into two and interchange the order of integration,

\begin{eqnarray}
\int\limits _{ 0 }^{ \infty  } dt 
\int\limits _{- t}^{ \infty } d \tau  
= 
\int\limits _{0}^{ \infty } d \tau  
\int\limits _{ 0 }^{ \infty  } dt 
+
\int\limits _{ - \infty  }^{ 0  } d \tau 
\int\limits _{- \tau }^{ \infty } d t  
\nonumber .
\end{eqnarray}
\ \\ \noindent
Then I have, 

\begin{eqnarray}
{\cal P}_{auto}^{ex}\left(  \omega \right) &=& R 
\frac{ e^{2} }{ {\cal T} _{0} }
 \eta_{}\left(   \omega \right)  ,
\label{ap102} \\
 \eta_{}\left(   \omega \right) &=&
\int\limits _{0}^{ \infty } d \tau  
\cos\left(  \omega  \tau  \right) 
e^{ -\frac{\tau }{ 2\tau_{D} }  }  
\frac{2 \sin\left( \omega_{0} \tau \right) }{ \pi \tau } 
.
\nonumber  
\end{eqnarray}
\ \\ \noindent
The suppression factor $ \eta( \omega)$ is shown in Fig.~\ref{fig4}. 

Let us first consider the case of $ \omega = 0$, when 

\begin{eqnarray}
 \eta_{}\left(  0 \right) &=&
\int\limits _{0}^{ \infty } d \tau  
 e^{ -\frac{\tau }{ 2\tau_{D} }  }  
\frac{ 2 \sin\left( \omega_{0} \tau \right) }{ \pi \tau } .
\label{ap103}
\end{eqnarray}
\noindent \\ 
The integrand in the above equation has two factors, exponentially decaying and oscillating. 
The wave function $ \psi^{ \Delta}$, Eq.~(\ref{06}), was calculated in the limit of $\omega_{0} \tau_{D} \gg 1$. \cite{Haack:2013ch,Moskalets:2013dl} 
Therefore, the period of oscillations is much smaller then the time of decay. 
Therefore, namely the fast oscillating factor determines an integral. 
The other, slowly decaying term can be calculated merely at $ \tau = 0$. 
So, by using a textbook integral 

$$
\int\limits _{0}^{ \infty } d \tau  
\frac{ \sin\left( \omega_{0} \tau \right) }{ \tau } = \frac{ \pi }{ 2 },
$$
I arrive at $ \eta\left(  0 \right) = 1$. 

For non zero frequencies, I represent the factor $ \eta\left(  \omega \right)$ as follows,

\begin{eqnarray}
 \eta_{}\left(   \omega \right) &=& 
\int\limits _{0}^{ \infty } d \tau  
e^{ -\frac{\tau }{ 2\tau_{D} }  }  
\frac{ \sin\left( \left[ \omega_{0} + \omega_{} \right] \tau \right) +  \sin\left( \left[ \omega_{0} - \omega_{} \right] \tau \right) }{ \pi \tau } .
\nonumber \\
\label{ap104} 
\end{eqnarray}
\ \\ \noindent
At small frequencies, $ \omega \ll \omega_{0}$, I can neglect $ \omega$ compared to $ \omega_{0}$, both sinuses becomes the same, and I recover $ \eta\left(  \omega \ll \omega_{0} \right) =  \eta(0)  = 1$. 
With this result, I reproduce Eq.~(\ref{07}), the first line.

At $ \omega = \omega_{0}$, the second sinus nullifies, and the result is halved, $ \eta\left(  \omega_{0} \right) = 0.5$. 
At higher frequencies, $ \omega > \omega_{0}$, the two sinuses contribute the same, but with opposite sign, and the noise gets suppressed, $ \eta\left(  \omega \right) = 0$. 
The transition from one to zero happens to occur within the region of order $ \tau_{D}^{-1}$ near $ \omega = \omega_{0}$ (that is, near $ \hbar \omega = \Delta/2$) in full agreement with the previous findings.\cite{Parmentier:2012ed,Moskalets:2013ed} 
Here I demonstrated that the wave function $ \psi^{ \Delta}(t)$, Eq.~(\ref{06}), carries information about this noise suppression effect.  

The calculations leading from Eq.~(\ref{01}), the second line with the wave function from Eq.~(\ref{06}), to Eq.~(\ref{07}), the second line, are rather straightforward. 
Importantly, the cross-correlation noise gets suppressed at much smaller frequencies of order $\tau_{D}^{-1} \ll \omega_{0}$.

\section{Relation between an electrical noise and an electron correlation function}
\label{ap02}

Here I generalize the results of Sec.~\ref{2c}. 
For this, it is convenient to separate the terms linear and bilinear in the Fermi functions $f_{1}$ and $f_{2}$ in Eqs.~(\ref{n06-1}). 
I will distinguish such terms via the upper index ${l}$ and ${b}$, respectively,

\begin{eqnarray}
{\cal P}_{33}\left(   \omega \right) &=& 
\frac{e^2}{h}
\int _{}^{ }
dE F_{33}(E,E+\hbar\omega) + 
{\cal P}_{33}^{l}\left(  \omega \right) + 
{\cal P}_{33}^{b}\left(   \omega \right) ,
\nonumber \\
\label{ap0200} \\
{\cal P}_{34}\left(   \omega \right) &=& {\cal P}_{34}^{l}\left(  \omega \right) + {\cal P}_{34}^{b}\left(  \omega \right).
\nonumber 
\end{eqnarray}
\noindent \\

\subsection{Auto-correlation nose}

Let us first consider ${\cal P}_{33}$. 
Substituting Eq.~(\ref{n05})  into Eq.~(\ref{n06-1}), I find 

\begin{eqnarray}
{\cal P}_{33}^{}(\omega) = 
\frac{e^2}{h}\int\limits_{}^{}dE
\Big\{
F_{33}(E,E+\hbar\omega)
+ 
\sum\limits_{n,m,q=- \infty}^{ \infty}
\nonumber \\
\label{ap0201} \\
R^{2}  f_{11}(E_q + \hbar\omega, E_{}) 
S_{F}^{*}(E_{n},E_{}) 
S_{F}(E_{m}, E_{})
\nonumber \\
\times
S_{F}^{*}(E_{m}+\hbar\omega,E_{q}+\hbar\omega)
S_{F}(E_{n}+\hbar\omega, E_q+\hbar\omega) 
\nonumber \\
+
RT f_{21}(E_q + \hbar\omega, E_{}) 
S_{F}^{*}(E_{n},E_{}) 
S_{F}(E_{m}, E_{})
\delta_{nq}\delta_{mq}
\nonumber \\
+
TR f_{12}(E_q + \hbar\omega, E_{}) 
\delta_{n0}\delta_{m0}
\nonumber \\
\times
S_{F}^{*}(E_{m}+\hbar\omega,E_{q}+\hbar\omega)
S_{F}(E_{n}+\hbar\omega, E_q+\hbar\omega) 
\nonumber \\
+ T^{2}f_{22}(E_q + \hbar\omega, E_{}) 
\delta_{n0}\delta_{m0}
\delta_{nq}\delta_{mq}
\Big\}
.
\nonumber 
\end{eqnarray}
\ \\ \noindent
Then, I use Eq.~(\ref{n07}) and calculate a linear in Fermi functions $f_{1}$ and $f_{2}$ part, 

\begin{eqnarray}
{\cal P}_{33}^{l}(\omega) &=& 
\frac{e^2}{2h}\int\limits_{}^{}dE \Big\{ 
Rf_{1}(E) + Rf_{1}(E+ \hbar \omega) +
\nonumber \\
\label{ap0202} \\
&&
 + Tf_{2}(E) + Tf_{2}(E+ \hbar \omega)  
\Big\} .
\nonumber 
\end{eqnarray}
\noindent \\
In the course of calculations, I utilized unitarity of the Floquet scattering matrix, see Eq.~(\ref{n04}). 
For a single-channel $S_{F}$, unitarity implies the following, 

\begin{eqnarray}
\sum\limits_{n=-\infty}^{\infty}
S_{F}^{*}\left(  E_{n}, E_{m} \right)
S_{F}^{}\left(  E_{n}, E_{} \right) = \delta_{m,0}, 
\label{ap0203} \\
\sum\limits_{n=-\infty}^{\infty}
S_{F}^{*}\left(  E_{m}, E_{n} \right)
S_{F}^{}\left(  E_{}, E_{n} \right) = \delta_{m,0}.  
\nonumber 
\end{eqnarray}
\noindent \\
As an example, let us consider the very first term we need to calculate, 

\begin{eqnarray}
\frac{e^2}{2 h}\int\limits_{}^{}dE
\sum\limits_{n,m,q=- \infty}^{ \infty}
R^{2}  f_{1}(E_q + \hbar\omega)  
S_{F}^{*}(E_{n},E_{}) 
S_{F}(E_{m}, E_{})
\nonumber \\
\times
S_{F}^{*}(E_{m}+\hbar\omega,E_{q}+\hbar\omega)
S_{F}(E_{n}+\hbar\omega, E_q+\hbar\omega) 
.
\nonumber 
\end{eqnarray}
\noindent \\
To simplify it, I shift $E_{q} \to E$ under integration over energy and shift $q \to -q$, $n-q \to n$, $m-q \to m$ under the corresponding sums, 

\begin{eqnarray}
\frac{e^2}{2 h}\int\limits_{}^{}dE
\sum\limits_{n,m,q=- \infty}^{ \infty}
R^{2}  f_{1}(E + \hbar\omega)  
S_{F}^{*}(E_{n},E_{q}) 
S_{F}(E_{m}, E_{q})
\nonumber \\
\times
S_{F}^{*}(E_{m}+\hbar\omega,E_{}+\hbar\omega)
S_{F}(E_{n}+\hbar\omega, E +\hbar\omega) 
.
\nonumber 
\end{eqnarray}
\noindent \\
Then, the sum over $q$ gives us, $ \sum_{q}^{}S_{F}^{*}(E_{n},E_{q}) 
S_{F}(E_{m}, E_{q}) = \delta_{n,m}$, which is used to sum over, say, $m$. 
The remaining sum over $n$ gives $ \sum_{n}^{} \left | S_{F}(E_{n}+\hbar\omega, E +\hbar\omega)   \right |^{2} = 1$. 
Therefore, what left is $\frac{e^2}{2 h}\int_{0}^{\infty}dE R^{2}  f_{1}(E + \hbar\omega)$. 
Other terms are calculated by analogy. 

The bilinear in Fermi functions $f_{1}$ and $f_{2}$ part reads, 

\begin{eqnarray}
{\cal P}_{33}^{b}(\omega) = - 
\frac{e^2}{h}\int\limits_{}^{}dE
\sum\limits_{n,m,q=- \infty}^{ \infty}
\Big\{
\nonumber \\
\label{ap0204} \\
R^{2}   f_{1}(E_q + \hbar\omega)  f_{1}(E_{})
S_{F}^{*}(E_{n},E_{}) 
S_{F}(E_{m}, E_{})
\nonumber \\
\times
S_{F}^{*}(E_{m}+\hbar\omega,E_{q}+\hbar\omega)
S_{F}(E_{n}+\hbar\omega, E_q+\hbar\omega) 
\nonumber \\
+
RTf_{2}(E_q + \hbar\omega) f_{1}(E_{})
S_{F}^{*}(E_{n},E_{}) 
S_{F}(E_{m}, E_{})
\delta_{nq}\delta_{mq}
\nonumber \\
+
TR f_{1}(E_q + \hbar\omega)  f_{2}(E_{})
\delta_{n0}\delta_{m0}
\nonumber \\
\times
S_{F}^{*}(E_{m}+\hbar\omega,E_{q}+\hbar\omega)
S_{F}(E_{n}+\hbar\omega, E_q+\hbar\omega) 
\nonumber \\
+ T^{2}f_{2}(E_q + \hbar\omega)  f_{2}(E_{})
\delta_{n0}\delta_{m0}
\delta_{nq}\delta_{mq}
\Big\}
.
\nonumber 
\end{eqnarray}
\ \\ \noindent
Let us represent ${\cal P}_{33}^{2}(\omega) = \sum_{j=1}^{4} B_{j}$ and calculate various terms separately. 
The first term is,  

\begin{eqnarray}
B_{1} &=& - 
\frac{e^2}{h}\int\limits_{}^{}dE
\sum\limits_{n,m,q=- \infty}^{ \infty}
\label{ap0205} \\
&&
R^{2}   f_{1}(E_q + \hbar\omega)  f_{1}(E_{})
S_{F}^{*}(E_{n},E_{}) 
S_{F}(E_{m}, E_{})
\nonumber \\
&&
\times
S_{F}^{*}(E_{m}+\hbar\omega,E_{q}+\hbar\omega)
S_{F}(E_{n}+\hbar\omega, E_q+\hbar\omega)
\nonumber \\
&=& - R^{2} v_{ \mu}^{2} e^{2} 
\int\limits _{ - {\cal T} _{0}/2 }^{ {\cal T} _{0}/2 } \frac{dt }{ {\cal T} _{0} }  \int\limits _{- \infty}^{ \infty } d \tau e^{i \omega \tau}  
\left | {\cal G}_{1}^{(1)}\left(  t + \tau; t \right)   \right |^{2}
,
\nonumber 
\end{eqnarray}
\noindent \\
where $ {\cal G}_{1}^{(1)}$ is the first-order correlation function of the Fermi sea incoming from the first channel and modified by the source, see Eq.~(\ref{cf01}). 
To prove the last line, I compute the time integral explicitly:  

\begin{eqnarray}
\int\limits _{ - {\cal T} _{0}/2 }^{ {\cal T} _{0}/2 } \frac{dt }{ {\cal T} _{0} }  \int\limits _{- \infty}^{ \infty } d \tau e^{i \omega \tau}  
\int \frac{dE }{ h } f_{ 1}\left( E \right) 
e^{- \frac{ i }{ \hbar  } E  \tau } 
\sum_{n,m=-\infty}^{\infty} 
e ^{-i \Omega  n \tau}
\nonumber \\
e ^{-i \Omega \left( n - m  \right) t}
S_{F}^{}\left(E_{n},E  \right)
S_{F}^{*}\left(E_{m},E  \right)
\int \frac{dE ^{\prime} }{ h } f_{ 1}\left( E ^{\prime} \right) 
e^{ \frac{ i }{ \hbar  } E ^{\prime}  \tau } 
\nonumber \\
\sum_{q, \ell=-\infty}^{\infty} 
e ^{i \Omega  q \tau}
e ^{i \Omega \left( q - \ell  \right) t}
S_{F}^{*}\left(E ^{\prime}_{q},E ^{\prime}  \right)
S_{F}^{}\left(E ^{\prime}_{ \ell},E ^{\prime}  \right) =
\nonumber 
\end{eqnarray}
\noindent \\
The integration over $t$ gives $ q-\ell = n-m$, which I use to sum up over $ \ell = q-n+m$ 

\begin{eqnarray}
= 
\int\limits _{- \infty}^{ \infty } d \tau e^{i \omega \tau}  
\int \frac{dE }{ h } f_{ 1}\left( E \right) 
e^{- \frac{ i }{ \hbar  } E  \tau } 
\sum_{n,m,q=-\infty}^{\infty} 
e ^{-i \Omega  n \tau}
\nonumber \\
S_{F}^{}\left(E_{n},E  \right)
S_{F}^{*}\left(E_{m},E  \right)
\int \frac{dE ^{\prime} }{ h } f_{ 1}\left( E ^{\prime} \right) 
e^{ \frac{ i }{ \hbar  } E ^{\prime}  \tau } 
e ^{i \Omega  q \tau}
\nonumber \\
S_{F}^{*}\left(E ^{\prime}_{q},E ^{\prime}  \right)
S_{F}^{}\left(E ^{\prime}_{ q-n+m},E ^{\prime}  \right) .
\nonumber 
\end{eqnarray}
\noindent \\
The integration over $ \tau$ gives $ h \delta\left( \hbar \omega - E_{n} + E_{q} ^{\prime}  \right)$, which I use to integrate out $E_{n} = E ^{\prime}_{q} + \hbar \omega$, 

\begin{eqnarray}
\sim
f_{ 1}\left( E ^{\prime}_{q-n} + \hbar \omega \right) 
f_{ 1}\left( E ^{\prime} \right) 
S_{F}^{*}\left(E ^{\prime}_{q},E ^{\prime}  \right)
S_{F}^{}\left(E ^{\prime}_{ q-n+m},E ^{\prime}  \right) 
\nonumber \\
S_{F}^{*}\left(E ^{\prime}_{q-n+m} + \hbar \omega,E ^{\prime}_{q-n} + \hbar \omega  \right)
S_{F}^{}\left(E ^{\prime}_{q} + \hbar \omega,E ^{\prime}_{q-n} + \hbar \omega  \right) .
\nonumber 
\end{eqnarray}
\noindent \\
Additionally I shift $q-n \to q$,

\begin{eqnarray}
\sim
f_{ 1}\left( E ^{\prime}_{q} + \hbar \omega \right) 
f_{ 1}\left( E ^{\prime} \right) 
S_{F}^{*}\left(E ^{\prime}_{q+n},E ^{\prime}  \right)
S_{F}^{}\left(E ^{\prime}_{ q+m},E ^{\prime}  \right) 
\nonumber \\
S_{F}^{*}\left(E ^{\prime}_{q+m} + \hbar \omega,E ^{\prime}_{q} + \hbar \omega  \right)
S_{F}^{}\left(E ^{\prime}_{q+n} + \hbar \omega,E ^{\prime}_{q} + \hbar \omega  \right) ,
\nonumber 
\end{eqnarray}
\noindent \\
and finally, I shift $q+n \to n$ and $q+m \to m$ and get the same integrand as in Eq.~(\ref{ap0205}) (up to $E ^{\prime} \to E$), 

\begin{eqnarray}
\sim
f_{ 1}\left( E ^{\prime}_{q} + \hbar \omega \right) 
f_{ 1}\left( E ^{\prime} \right) 
S_{F}^{*}\left(E ^{\prime}_{n},E ^{\prime}  \right)
S_{F}^{}\left(E ^{\prime}_{ m},E ^{\prime}  \right) 
\nonumber \\
S_{F}^{*}\left(E ^{\prime}_{m} + \hbar \omega,E ^{\prime}_{q} + \hbar \omega  \right)
S_{F}^{}\left(E ^{\prime}_{n} + \hbar \omega,E ^{\prime}_{q} + \hbar \omega  \right) .
\nonumber 
\end{eqnarray}
\noindent \\
Other terms are calculated in the same way,
 
\begin{eqnarray}
B_{2} = - 
\frac{e^2}{h}\int\limits_{}^{}dE
\sum\limits_{q= - \infty}^{ \infty}
RTf_{2}(E_q + \hbar\omega) f_{1}(E_{})
\nonumber \\
\times
\left | S_{F}(E_{q}, E_{}) \right |^{2}
= - RT v_{ \mu}^{2} e^{2} 
\quad 
\label{ap0206} \\
\times
\int\limits _{ - {\cal T} _{0}/2 }^{ {\cal T} _{0}/2 } \frac{dt }{ {\cal T} _{0} }  \int\limits _{- \infty}^{ \infty } d \tau e^{i \omega \tau}  
{\cal G}_{1}^{(1)}\left(  t + \tau; t \right)  
{\cal G}_{0,2}^{(1)*}\left(   \tau  \right)  
,
\nonumber 
\end{eqnarray}

\begin{eqnarray}
B_{3} = - 
\frac{e^2}{h}\int\limits_{}^{}dE
\sum\limits_{q= - \infty}^{ \infty}
TR f_{1}(E_q + \hbar\omega)  f_{2}(E_{})
\nonumber \\
 \times
\left | S_{F}(E_{}+\hbar\omega, E_q+\hbar\omega) \right |^{2} 
= - TR v_{ \mu}^{2} e^{2} 
\quad 
\label{ap0207} \\
\times
\int\limits _{ - {\cal T} _{0}/2 }^{ {\cal T} _{0}/2 } \frac{dt }{ {\cal T} _{0} }  \int\limits _{- \infty}^{ \infty } d \tau e^{i \omega \tau}  
{\cal G}_{0,2}^{(1)}\left(  \tau  \right)  
{\cal G}_{1}^{(1)*}\left(  t + \tau; t \right)  
,
\nonumber 
\end{eqnarray}

\begin{eqnarray}
B_{4} &=& - 
\frac{e^2}{h}\int\limits_{}^{}dE
T^{2}f_{2}(E + \hbar\omega)  f_{2}(E_{})
\label{ap0208} \\
&=& - T^{2} v_{ \mu}^{2} e^{2} 
\int\limits _{ - {\cal T} _{0}/2 }^{ {\cal T} _{0}/2 } \frac{dt }{ {\cal T} _{0} }  \int\limits _{- \infty}^{ \infty } d \tau e^{i \omega \tau} \left | {\cal G}_{0,2}^{(1)}\left(  \tau  \right) \right |^{2} .
\nonumber 
\end{eqnarray}
\ \\ \noindent
Combing all terms together I get, 

\begin{eqnarray}
{\cal P}_{ 33}^{b}\left(  \omega \right)  &=& 
- v_{ \mu}^{2} e^{2} 
 \int\limits _{ - {\cal T} _{0}/2 }^{ {\cal T} _{0}/2 } \frac{dt }{ {\cal T} _{0} }  
\int\limits _{- \infty}^{ \infty } d \tau e^{i \omega \tau}  
\nonumber \\
\label{ap0209} \\
&&
\left |  R {\cal G}_{1}^{(1)}\left(  t + \tau; t \right)  + T {\cal G}_{0,2}^{(1)}\left(  \tau  \right)   \right |^{2} .
\nonumber 
\end{eqnarray}
\noindent \\
Note that when the second channel is also fed by a source, we need to replace the correlation function of the Fermi sea in equilibrium, ${\cal G}_{0,2}^{(1)}$, by the one modified by the source, ${\cal G}_{2}^{(1)}$. 

Strictly speaking, the total auto-correlation noise, ${\cal P}_{33}^{}(\omega)$,  Eq.~(\ref{ap0200}), diverges. 
This is due to the infinite number of electrons that make up the Fermi sea, which contribute to both ${\cal P}_{33}^{l}(\omega)$, Eq.~(\ref{ap0202}),  and ${\cal P}_{33}^{b}(\omega)$, Eq.~(\ref{ap0209}).    
Therefore, to see the effect of injected electrons, we need to look at the excess noise. 
For this, I first calculate the noise when the source is off, ${\cal P}_{33}^{off}$, and then calculate the excess noise as the difference, ${\cal P}_{33}^{ex} = {\cal P}_{33}^{} - {\cal P}_{33}^{off}$.

\subsubsection{Excess noise}

For the sake of simplicity, I assume all the reservoirs have the same Fermi energy, $ \mu_{ \alpha} = \mu, \forall \alpha$. 
If necessary, the constant bias $V_{ \alpha}$ at lead $ \alpha$, that is, $ \mu_{ \alpha} = \mu + e V_{ \alpha}$, can be accounted for via an energy-independent scattering amplitude $S_{ \alpha} = e^{i \frac{ e V_{ \alpha} }{ \hbar  } t}$. 

To calculate ${\cal P}_{33}^{off}$, I replace ${\cal G}_{1}$ by ${\cal G}_{0,1}$ in Eq.~(\ref{ap0209}), use Eq.~(\ref{ap0202}), which remains unchanged, and get according to Eq.~(\ref{ap0200}), 

\begin{eqnarray}
{\cal P}_{33}^{off}\left(  \omega \right) &=& 
\frac{e^2}{h}\int\limits_{}^{}dE 
\Bigg\{ 
F_{33}(E,E+\hbar\omega) 
+ 
\label{ap0210} \\
&&
R^{2} F_{11}(E + \hbar\omega, E) 
+ TR F_{21}(E + \hbar\omega, E) 
\nonumber \\ &&
+ RT F_{12}(E + \hbar\omega, E) 
+ T^{2} f_{22}(E + \hbar\omega, E) 
\Bigg\} .
\nonumber 
\end{eqnarray}
\noindent \\
Then I transform,  

\begin{eqnarray}
F_{ \alpha \beta}(E,E+\hbar\omega) + F_{  \beta \alpha}(E,E+\hbar\omega) = 
\nonumber \\
= F_{ \alpha \alpha}(E,E+\hbar\omega) + 
F_{ \beta \beta}(E,E+\hbar\omega) + 
\Phi_{ \alpha \beta} ,
\nonumber \\
\Phi_{ \alpha \beta} = 
\left[  f_{ \alpha}\left(  E \right) - f_{ \beta}\left(  E \right)\right]
\left[  f_{ \alpha}\left(  E + \hbar \omega \right) - f_{ \beta}\left(  E + \hbar \omega \right) \right] ,
\label{ap0210x}
\nonumber 
\end{eqnarray}
\noindent \\
and integrate over energy,

\begin{eqnarray}
\int\limits_{}^{}dE 
F_{ \alpha \alpha}(E,E+\hbar\omega) = \frac{ \hbar \omega }{ 2  } \coth \frac{ \hbar \omega }{ 2 k_{B} \theta_{ \alpha} } ,
\nonumber \\
\label{ap0210xx} \\
\Xi \left(  \theta_{ \alpha}, \theta_{ \beta} , \omega \right) = \frac{ 1 }{ k_{B} \left(  \theta_{ \alpha} + \theta_{ \beta}  \right) }
\int\limits_{}^{} dE \Phi_{ \alpha \beta}   = 
\frac{ 2  \theta_{ \alpha} }{  \theta_{ \alpha} + \theta_{ \beta}  } 
\nonumber \\
\times
\int\limits _{0}^{ 1 } dx 
\frac{ 
x 
\left(  \left[ x \Omega \right]^{ \frac{ \theta_{ \alpha} }{ \theta_{ \beta}   } -1 } - 1 \right) 
\left(  \left[ \frac{ x }{ \Omega  } \right]^{ \frac{ \theta_{ \alpha} }{ \theta_{ \beta}   } -1 } - 1 \right)
}{ 
\left(  1 + x \Omega \right)  
\left(  1 + \left[ x \Omega \right]^{ \frac{ \theta_{ \alpha} }{ \theta_{ \beta}  }}  \right)
\left(   1 + \frac{ x }{ \Omega  } \right)
\left(   1 + \left[  \frac{ x }{ \Omega  } \right]^{ \frac{ \theta_{ \alpha} }{ \theta_{ \beta}  }} 
\right)
}
,
\nonumber 
\end{eqnarray}
\noindent \\
where $ \Omega = e^{ - \frac{ \hbar \omega }{ 2 k_{B} \theta_{ \alpha}  }}$ and $k_{B}$ is the Boltzmann constant.
Using the above equation, I represent ${\cal P}_{33}^{off}$ as follows, 

\begin{eqnarray}
{\cal P}_{33}^{off}\left(  \omega \right) = 
\frac{ e^{2} }{ h  } k_{B} \Bigg\{
\theta_{3} \xi_{ }\left(  \frac{ \hbar \omega }{ 2 k_{B} \theta_{3}  } \right)
+ R^{} \theta_{1} \xi_{ }\left(  \frac{ \hbar \omega }{ 2 k_{B} \theta_{1}  } \right)  
\nonumber  \\
+ T^{}   \theta_{2} \xi_{ }\left(  \frac{ \hbar \omega }{ 2 k_{B} \theta_{2}  } \right) 
+ RT \left(  \theta_{1} + \theta_{2} \right)  \Xi \left(  \theta_{ 1}, \theta_{ 2} , \omega \right)  
\Bigg\} , \quad
\label{ap0210-2} 
\end{eqnarray}
\noindent \\
where $ \xi_{ }\left( x \right) =  x \coth x$, which decreases monotonically  from one at $x=0$ to zero at $x \to \infty$. 

\begin{figure}[t]
\centering
\resizebox{0.99\columnwidth}{!}{\includegraphics{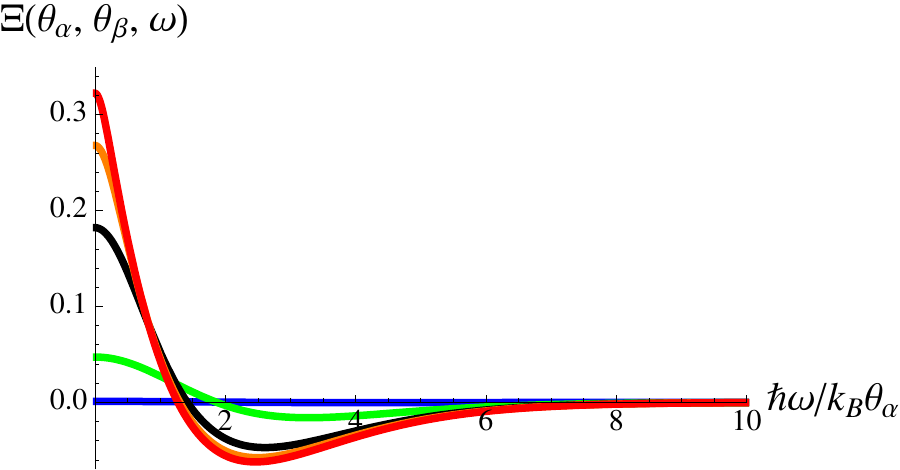} }
\caption{
The joint two terminal contribution to the thermal noise $ \Xi \left(  \theta_{ \alpha}, \theta_{ \beta} , \omega \right)$, see Eqs.~(\ref{ap0210xx}) and (\ref{ap0210-2}), is shown as a function of frequency at $ \theta_{ \alpha}/ \theta_{ \beta} = 1.1 \, ({\rm blue})$, $2 \, ({\rm green})$, $5 \, ({\rm black})$, $10 \, ({\rm orange})$, and  $20 \, ({\rm red})$.  
}
\label{fig5}
\end{figure}

The term with factor $ \Xi \left(  \theta_{ 1}, \theta_{ 2} , \omega \right)$ describes the auto-correlation thermal noise due to both terminals.
It is shown in Fig.~\ref{fig5}. 
Note that $ \Xi \left(  \theta_{ \alpha}, \theta_{ \beta} , \omega \right) $ is zero at $ \theta_{ \alpha} = \theta_{ \beta}$, and as a function of $ \omega$ it changes a sign from positive to negative with increasing $ \omega$. 
This sign change occurs around $ \hbar \omega = k_{B}\left(  \theta_{ \alpha} + \theta_{ \beta} \right)$. 
At this temperature/frequency, the joint two terminal contribution to thermal noise vanishes.

Note that while expressing Eq.~(\ref{ap0209}) in terms of $F_{ \alpha \beta}$, Eq.~(\ref{n07}),  I used,

\begin{eqnarray}
\frac{1}{h}\int\limits_{}^{}dE
\left\{ F_{ \alpha \beta}(E,E+\hbar\omega) - \frac{ f_{ \alpha}(E) + f_{ \beta}(E + \hbar \omega)  }{ 2 } \right\} = 
\nonumber \\
\label{ap0210-1} \\
= - v_{ \mu}^{2} 
\int\limits _{ - {\cal T} _{0}/2 }^{ {\cal T} _{0}/2 } \frac{dt }{ {\cal T} _{0} }  \int\limits _{- \infty}^{ \infty } d \tau e^{i \omega \tau}  {\cal G}_{0, \alpha}^{(1)}\left(   \tau  \right)  {\cal G}_{0, \beta}^{(1)*}\left(  \tau  \right) .
\nonumber 
\end{eqnarray}
\noindent \\
For examples, see Eq.~(\ref{ap0208}) and alike. 

The excess auto-correlation noise reads, 

\begin{eqnarray}
{\cal P}_{ 33}^{ex}\left(  \omega \right)  &=& 
e^{2} v_{ \mu}^{2} 
 \int\limits _{ - {\cal T} _{0}/2 }^{ {\cal T} _{0}/2 } \frac{dt }{ {\cal T} _{0} }  
\int\limits _{- \infty}^{ \infty } d \tau e^{i \omega \tau}  
\Bigg\{
\label{ap0211} \\
&&
\left |  R {\cal G}_{0,1}^{(1)}\left(  t + \tau; t \right)  + T {\cal G}_{0,2}^{(1)}\left(  \tau  \right)   \right |^{2} 
\nonumber \\
&&
-
\left |  R {\cal G}_{1}^{(1)}\left(  t + \tau; t \right)  + T {\cal G}_{0,2}^{(1)}\left(  \tau  \right)   \right |^{2} 
\Bigg\}.
\nonumber 
\end{eqnarray}
\noindent \\
When both incoming channels are at the same temperature, $ \theta_{1} = \theta_{2} \equiv \theta$, and have the same Fermi energies, $ \mu_{1} = \mu_{2} \equiv \mu$, that is, their equilibrium correlation functions are the same, ${\cal G}_{0,1} = {\cal G}_{0,2} \equiv {\cal G}_{0}$, the above equation is simplified, [see Eq.~(\ref{npg01a})]  

\begin{eqnarray}
{\cal P}_{ 33}^{ex}\left(  \omega \right)  &=& 
e^{2}  v_{ \mu}^{2} 
 \int\limits _{ - {\cal T} _{0}/2 }^{ {\cal T} _{0}/2 } \frac{dt }{ {\cal T} _{0} }  
\int\limits _{- \infty}^{ \infty } d \tau e^{i \omega \tau}  
\Bigg\{
\label{ap0211-1} \\
&&
- R^{2} \left | { G}_{1}^{(1)}\left(  t + \tau; t \right) \right |^{2} 
\nonumber \\
&&
- 2 R \,  {\rm Re} { G}_{1}^{(1)}\left(  t + \tau; t \right)  {\cal G}_{0}^{(1)*}\left(  \tau  \right) 
\Bigg\},
\nonumber 
\end{eqnarray}
\noindent \\ 
where 

\begin{eqnarray}
G ^{(1)}_{1}\left(  t + \tau; t \right) =  {\cal  G}_{1}^{(1)}\left(  t + \tau; t \right)   -  {\cal G}_{0,1}^{(1)}\left(  \tau  \right) . 
\label{ap0212}
\end{eqnarray}
\noindent \\ 
is  the excess first-order correlation function injected by the source,

\subsection{Cross-correlation nose}

The cross-correlation noise in terms of ${\cal G} ^{(1)}_{}$ was given in Ref.~\onlinecite{Moskalets:2015ub}, 

\begin{eqnarray}
{\cal P}_{34}(\omega)  &=& 
- RT e^{2} v_{ \mu}^{2} 
 \int\limits _{ - {\cal T} _{0}/2 }^{ {\cal T} _{0}/2 } \frac{dt }{ {\cal T} _{0} }  
\int\limits _{- \infty}^{ \infty } d \tau e^{i \omega \tau}  
\label{ap0213} \\
&& \times
\left | {\cal G}_{1}^{(1)}\left(  t + \tau; t \right)  - {\cal G}_{0,2}^{(1)}\left(  \tau   \right)   \right |^{2} .
\nonumber 
\end{eqnarray}
\noindent \\
The excess cross-correlation noise reads, 

\begin{eqnarray}
{\cal P}_{ 34}^{ex}\left(  \omega \right)  &=& 
RT e^{2} v_{ \mu}^{2} 
 \int\limits _{ - {\cal T} _{0}/2 }^{ {\cal T} _{0}/2 } \frac{dt }{ {\cal T} _{0} }  
\int\limits _{- \infty}^{ \infty } d \tau e^{i \omega \tau}  
\Bigg\{
\label{ap0214} \\
&&
\left | {\cal G}_{0,1}^{(1)}\left(   \tau  \right)  -  {\cal G}_{0,2}^{(1)}\left(  \tau  \right)   \right |^{2} 
\nonumber \\
&&
-
\left |   {\cal G}_{1}^{(1)}\left(  t + \tau; t \right)  - {\cal G}_{0,2}^{(1)}\left(  \tau  \right)   \right |^{2} 
\Bigg\}.
\nonumber 
\end{eqnarray}
\noindent \\
If both incoming channels are under the same conditions, ${\cal G}_{0,1} = {\cal G}_{0,2} \equiv {\cal G}_{0}$, then, unlike the auto-correlation noise,  the cross-correlation noise has no equilibrium contribution and it is expressed solely in terms of the excess correlation function $G ^{(1)}_{}$, [see Eq.~(\ref{npg01b})]

\begin{eqnarray}
{\cal P}_{34}(\omega)  &=& {\cal P}_{34}^{ex}(\omega) = 
- RT e^{2} v_{ \mu}^{2}  
 \int\limits _{ - {\cal T} _{0}/2 }^{ {\cal T} _{0}/2 } \frac{dt }{ {\cal T} _{0} }  
\int\limits _{- \infty}^{ \infty } d \tau e^{i \omega \tau}  
\nonumber \\
&& \times
\left |  { G}_{1}^{(1)}\left(  t + \tau; t \right)   \right |^{2} .
\label{ap0215}
\end{eqnarray}
\noindent \\

\subsection{The excess noise conservation at zero frequency}
\label{nc}

Conservation of charge imposes strong restrictions on zero frequency noise. \cite{Blanter:2000wi} 
Namely, in the case under consideration, ${\cal P}_{33}^{ex}\left(  0 \right)$ and  ${\cal P}_{34}^{ex}\left(  0 \right)$ are perfectly anti-correlated, such that 

\begin{eqnarray}
{\cal P}_{33}^{ex}\left(  0 \right) + {\cal P}_{34}^{ex}\left(  0 \right) = 0. 
\label{npg03}
\end{eqnarray}
\ \\ \noindent
To show this, let us use Eqs.~(\ref{ap0211}) and (\ref{ap0214}) and obtain,

\begin{eqnarray}
\int\limits _{ - {\cal T} _{0}/2 }^{ {\cal T} _{0}/2 } \frac{dt }{ {\cal T} _{0} }  
\int\limits _{- \infty}^{ \infty } d \tau \left |  {\cal G}_{1}^{(1)}\left(  t + \tau; t \right) \right |^{2} 
=
\label{npg04} \\
= \int\limits _{ - {\cal T} _{0}/2 }^{ {\cal T} _{0}/2 } \frac{dt }{ {\cal T} _{0} }  
\int\limits _{- \infty}^{ \infty } d \tau \left |  {\cal G}_{0,1}^{(1)}\left(   \tau   \right) \right |^{2} .
\nonumber 
\end{eqnarray}
\noindent \\
Then I use Eqs.~(\ref{cf01-1}), (\ref{cf01}) and calculate the right hand side in the above equation,  

\begin{eqnarray}
r.h.s. = \frac{ 1 }{ h v_{ \mu}^{2}  } \int _{}^{ } dE f_{1}^{2}(E) , 
\label{npg05}
\end{eqnarray}
\noindent \\ 
and the left hand side,  

\begin{eqnarray}
l.h.s. = \frac{ 1 }{ h v_{ \mu}^{2}  } \int _{}^{ } dE
\sum_{n,m,q}^{} 
f_{ 1}\left( E ^{}_{q} \right) 
f_{1 }\left( E ^{} \right) 
\quad 
\label{npg06} \\
\times
S_{F}^{*}\left(E ^{}_{n},E ^{}  \right)
S_{F}^{}\left(E ^{}_{ m},E ^{}  \right)
S_{F}^{*}\left(E ^{}_{m} ,E ^{}_{q}   \right)
S_{F}^{}\left(E ^{}_{n} ,E ^{}_{q}  \right) .
\nonumber
\end{eqnarray}
\noindent \\ 
Using the unitarity of the Floquet scattering amplitude of the source in the above equation, see Eqs.~(\ref{n04}) and (\ref{ap0203}), I sum over $n$ and $m$ and get the following,

\begin{eqnarray}
l.h.s. = \frac{ 1 }{ h v_{ \mu}^{2}  } \int _{}^{ } dE f_{1}^{2}(E) = r.h.s. \,, 
\label{npg07}
\end{eqnarray}
\noindent \\ 
that is, Eq.~(\ref{npg04}) is proven.

\subsection{Noise and the outgoing correlation matrix of a linear electronic circuit}
\label{ap02-4}

To generalize equations that relate noise to electron correlation functions to the case with several sources and/or with several outgoing channels, it is instructive to combine correlation functions of all incoming channels into a square matrix $\hat {\cal G} ^{(1)}_{in}$, and of all outgoing channels into a square matrix  $\hat {\cal G} ^{(1)}_{out}$, see Ref.~\onlinecite{Moskalets:2016fm} for details.   
Their dimensions are equal to the number of incoming and outgoing channels, respectively. 
Let us denote by $\hat S$ the scattering matrix of a stationary electronic circuit connecting incoming and outgoing channels. 
If $\hat S$ is independent of energy, the incoming and outgoing correlation matrix are related as follows, $\hat {\cal G} ^{(1)}_{out}\left(  t_{1}; t_{2} \right) = \hat S^{*} \hat {\cal G} ^{(1)}_{in}\left(  t_{1}; t_{2} \right) \hat S^{T}$, where the upper index ${*}$ means complex conjugation, and  ${T}$ means transposition. 
Since the incoming channels are not correlated, $\hat {\cal G} ^{(1)}_{in}$ is diagonal.  

In the case of a single QPC considered in this work, ${\cal G} ^{(1)}_{in,11} = {\cal G} ^{(1)}_{1}$, ${\cal G} ^{(1)}_{in,22} = {\cal G} ^{(1)}_{2}$, and two other elements are zero.
The circuit's scattering matrix is given in Eq.~(\ref{04n-1}), $\hat S = \hat S^{QPC}$. 
Then, the outgoing correlation matrix reads,

\begin{eqnarray}
\hat {\cal G} ^{(1)}_{out} =
\left(
\begin{array}{ccc}
R {\cal G} ^{(1)}_{1} + T {\cal G} ^{(1)}_{2}  & & i \sqrt{RT} \left(  {\cal G} ^{(1)}_{1} - {\cal G} ^{(1)}_{2} \right) \\
i \sqrt{RT} \left(  {\cal G} ^{(1)}_{2} - {\cal G} ^{(1)}_{1} \right)  & &  T {\cal G} ^{(1)}_{1} + R {\cal G} ^{(1)}_{2}  
\end{array}
\right).
\nonumber \\
\label{ap02out}
\end{eqnarray}
\noindent \\
Comparing the elements of $\hat {\cal G} ^{(1)}_{out}$ to Eqs.~(\ref{ap0211}) and (\ref{ap0214}), we can relate the excess noise and the elements of the outgoing correlation matrix as follows, 

\begin{eqnarray}
{\cal P}_{ \alpha \alpha ^{\prime}}^{ex}\left(  \omega \right)  =
e^{2}  
 \int\limits _{ - {\cal T} _{0}/2 }^{ {\cal T} _{0}/2 } \frac{dt }{ {\cal T} _{0} }  
\int\limits _{- \infty}^{ \infty } d \tau e^{i \omega \tau}  
\quad
\label{ap02ng} \\
\left\{ 
\left | v_{ \mu} {\cal G}^{(1), off}_{out, \alpha \alpha ^{\prime}} \left(  t + \tau; t \right)  \right |^{2} 
- 
\left | v_{ \mu} {\cal G}^{(1),on}_{out, \alpha \alpha ^{\prime}} \left(  t + \tau; t \right)  \right |^{2} 
\right\} ,
\nonumber 
\end{eqnarray}
\noindent \\
where the upper indices ${off}$ and ${on}$ indicate that electronic sources are off and on, respectively. 

Notice that in the above equation, the indices $ \alpha$ and $ \alpha ^{\prime}$ number outgoing channels only, therefore, $ \alpha, \alpha ^{\prime} = 1,2$. 
While in Eqs.~(\ref{ap0211}) and (\ref{ap0214}), the outgoing channels were numbered together with incoming channels, hence the outgoing channels were $3$ and $4$, respectively.  
Equation (\ref{ap02ng}) is applicable for any number of electronic sources and any number of outgoing channels.

\section{Noise caused by an electron injected from a quantum level raising at a constant rapidity at zero temperature}
\label{ap03}

In this appendix I derive Eqs.~(\ref{c01}) starting from Eqs.~(\ref{npg01}). 

First, I use the wave function $ \psi^{(c)}$ from Eq.~(\ref{04}), and calculate 
the excess correlation function,

\begin{eqnarray}
G ^{(1)}\left( t + \tau; t_{} \right) &=&
 \frac{ e ^{ i \frac{ \mu }{ \hbar } \tau}  }{ \pi  \Gamma _{\tau} v_{ \mu} }
\int\limits_{0}^{ \infty} dx 
e^{ -  x } 
e^{   i x \frac{ t + \tau   }{  \Gamma _{\tau}  }     } 
e^{ -i x^{2} \zeta  }  
\nonumber \\
&&
\int\limits_{0}^{ \infty} dy 
e^{ -  y } 
e^{  - i y \frac{ t_{}   }{  \Gamma _{\tau}  }     } 
e^{ i y^{2} \zeta  } .
\label{ap0301} 
\end{eqnarray}
\ \\ \noindent
Here I introduced the parameter of non-adiabaticity $ \zeta = \tau_{D}/ \Gamma _{\tau}$. 
This parameter controls the symmetry of the density profile of the injected wave packet. \cite{Keeling:2008ft}
With a symmetric density profile, injection is classified as adiabatic, with an asymmetric density profile, injection is non-adiabatic.\cite{Moskalets:2013dl}
At $ \zeta = 0$, this source is identical to the source of levitons, that is, $ \psi^{c} = \psi^{L}$, see Eqs.~(\ref{02}) and (\ref{04}). 

At the next step, let us calculate separately two terms entering Eqs.~(\ref{npg01}). 
I assume $ \Gamma _{\tau} \ll {\cal T} _{0}$ and, therefore, extend the limits of integration over $t$ to infinity. 

\subsection{ The term with $- v_{ \mu}^{2} G_{1} ^{(1)}  {\cal G}_{0} ^{(1)*}$}
\label{ap3s1}

I denote this term as ${\Pi}_{1}$. 
Using Eq.~(\ref{cf01-1}) for ${\cal G}_{0} ^{(1)}$ at zero temperature, I have, 

\begin{eqnarray}
{\Pi}_{1}^{}(\omega) &=& - 
\int\limits_{- \infty}^{ \infty } d \tau 
e^{i \omega \tau }
2{\rm Re}
\frac{ {\cal C}\left(   \tau \right) }{ - 2 \pi i \tau  } ,
\label{ap0302} \\
\nonumber \\
{\cal C}\left(   \tau \right) &=&
\frac{ 1  }{ \pi  \Gamma _{\tau}  }
\int\limits _{ - \infty }^{ \infty } dt 
\int\limits_{0}^{ \infty} dx 
e^{ -  x } 
e^{   i x \frac{ t + \tau  }{  \Gamma _{\tau}  }     } 
e^{ -i x^{2} \zeta  }  
\nonumber \\
&&
\times
\int\limits_{0}^{ \infty} dy 
e^{ -  y } 
e^{  - i y \frac{ t_{}   }{  \Gamma _{\tau}  }     } 
e^{ i y^{2} \zeta  } ,
\nonumber  
\end{eqnarray}
\ \\ \noindent
First, I evaluate coherence ${\cal C}\left(  \tau \right)$. 
For this, I integrate over $t$, 

$$
\int\limits _{ - \infty }^{ \infty } dt   
e^{   i x \frac{ t  }{  \Gamma _{\tau}  }     } 
e^{  - i y \frac{ t_{}   }{  \Gamma _{\tau}  }     } = 2 \pi \Gamma _{\tau}  \delta\left( x - y \right).
$$
This allows us to integrate, say, over $y$ (the result is independent of the  parameter of non-adiabaticity $ \zeta$), 

\begin{eqnarray}
{\cal C}\left(  \tau \right)  = 
2 
\int\limits_{0}^{ \infty}  dx 
e^{ -  2x } 
e^{   i  \frac{ x  }{  \Gamma _{\tau}  }  \tau   } 
= 
\frac{ 1 }{ 1 - i \tau/\left(  2 \Gamma _{\tau} \right) }
.
\label{ap0303} 
\end{eqnarray}
\ \\ \noindent
Substituting the above equation into Eq.~(\ref{ap0302}), I calculate 

\begin{eqnarray}
{\Pi}_{1}^{}(\omega) &=&  
e^{- \left | \omega \right | 2 \Gamma _{\tau}}
.
\label{ap0305} 
\end{eqnarray}
\ \\ \noindent
Note that namely this term determines the auto-correlation noise at $R \ll 1$: ${\cal P}_{ 33}^{ex}\left(  \omega \right) \approx R \left(  e^{2} / {\cal T} _{0} \right) {\Pi}_{1}^{}(\omega)$, see Eq.~(\ref{05}).

\subsection{ The term with  $ v_{ \mu}^{2} \left | G ^{(1)} \right |^{2}$}
\label{ap3s2}

I denote this term as ${\Pi}_{2}$,  

\begin{eqnarray}
{\Pi}_{2}(\omega)  = 
\int\limits _{ - \infty }^{ \infty} dt  
\int\limits_{- \infty}^{ \infty } d \tau 
e^{i \omega \tau }
\left |  \psi^{(c)}\left(  t + \tau \right) \right |^{2}
\left |  \psi^{(c)}\left(  t  \right) \right |^{2} .
\label{ap0306} 
\end{eqnarray}
\ \\ \noindent
First, I shift $ \tau + t \to \tau$, and represent the above equation as the  square of the Fourier transform of the wave packet density, ${\Pi}_{2}(\omega) = \left | {\cal N} \left(   \omega \right) \right |^{2}$, where

\begin{eqnarray}
 {\cal N}\left(  \omega \right) &=& \Gamma _{\tau}
\int\limits_{- \infty}^{ \infty } d z 
e^{i \omega  \Gamma _{\tau} z } \left | \psi^{(c)}\left(  \Gamma _{\tau} z \right)  \right |^{2} = 
\frac{1 }{ \pi^{}   }
\int\limits_{- \infty}^{ \infty } d z 
e^{i \omega  \Gamma _{\tau} z }
\nonumber \\
\label{ap0307} \\
&&
\times
\int\limits_{0}^{ \infty} dx 
e^{ -  x } 
e^{   i x   z     } 
e^{ -i x^{2} \zeta  }  
\int\limits_{0}^{ \infty} dy 
e^{ -  y } 
e^{  - i y  z        } 
e^{ i y^{2} \zeta  }  ,
\nonumber 
\end{eqnarray}
\ \\ \noindent
where $z = \tau/  \Gamma _{\tau}$.
The integration over $ z$ gives, 

$$
\int\limits_{- \infty}^{ \infty } d z  
e^{i \omega  \Gamma _{\tau} z }
e^{   i x   z     } 
e^{  - i y  z        } 
=  2 \pi \delta\left(  \omega  \Gamma _{\tau} + x - y \right) .
$$
Then I integrate over $y = x + \omega  \Gamma _{\tau}$ (for positive $ \omega$), 

\begin{eqnarray}
 {\cal N}\left(  \omega \right) &=& 
2
e^{ i \left( \omega  \Gamma _{\tau} \right)^{2}  \zeta  }  
e^{ -  \omega  \Gamma _{\tau} } 
\int\limits_{0}^{ \infty}  
dx 
e^{ -  2x  } 
e^{ i 2 x \omega  \Gamma _{\tau} \zeta  } 
\nonumber \\
&=&
e^{ i \left( \omega  \Gamma _{\tau} \right)^{2}  \zeta  }  
\frac{ e^{ -  \omega  \Gamma _{\tau} }  }{  1 - i \omega  \Gamma _{\tau} \zeta } ,
\nonumber 
\end{eqnarray}
\ \\ \noindent
and obtain (for arbitrary $ \omega $), 

\begin{eqnarray}
{\Pi}_{2}(\omega) &=&
\frac{ e^{ -  \left | \omega \right | 2  \Gamma _{\tau} }  }{ 1 + \left( \omega  \tau_{D} \right)^{2} } ,
\label{ap0308}
\end{eqnarray}
\ \\ \noindent
where $ \tau_{D} = \Gamma _{\tau} \zeta$. 
This term fully defines cross-correlation noise and partially auto-correlation noise.

\subsection{ Excess noise power}

By combining both terms together, I get,

\begin{eqnarray}
{\cal P}_{33}^{ex}(\omega) &=& 
R \frac{ e^{2} }{   {\cal T} _{0} } {\Pi}_{1}(\omega)
- 
R^{2} \frac{ e^{2} }{   {\cal T} _{0} } {\Pi}_{2}(\omega)
\nonumber \\
\label{ap0309} \\
{\cal P}_{34}^{ex}(\omega) &=& 
- R T \frac{ e^{2} }{   {\cal T} _{0} } {\Pi}_{2}(\omega) .
\nonumber 
\end{eqnarray}
\noindent \\
Using Eqs.~(\ref{ap0305}) and (\ref{ap0308}) we arrive at Eq.~(\ref{c01}). 

Note that at $ \omega = 0$, ${\Pi}_{1}(0) = {\Pi}_{2}(0)$ and $ {\cal P}_{33}^{ex}(0) +  {\cal P}_{34}^{ex}(0) = 0$ in agreement with Eq.~(\ref{npg03}). 

\section{Noise caused by an electron injected from a quantum level raising at a constant rapidity at non-zero temperature}
\label{ap04}

In this appendix, I repeat the calculations presented in the previous Appendix, but at non-zero temperatures, and compute the temperature-dependent factor $ \eta_{ }\left(   \omega, \theta \right) $ from Eq.~(\ref{c02}). 

For this model, the correlation function of electrons injected at a nonzero temperature, $ \theta > 0$, was expressed in terms of the correlation function at zero temperature in Ref.~\onlinecite{Moskalets:2017fh}, 

\begin{eqnarray}
G ^{(1)}_{1, \theta} \left(  t + \tau; t \right) &=& 
\int _{}^{ } d \epsilon 
\left(  - \frac{  \partial f_{1} }{  \partial \epsilon  } \right)
\nonumber \\
&& \times
e^{i \frac{ \epsilon }{ \hbar } \tau }
G ^{(1)}_{1, 0} \left(  t + \tau - \frac{ \epsilon }{ c  }; t - \frac{ \epsilon }{ c  } \right). 
\label{t01}
\end{eqnarray}
\noindent \\
Here the indices $ \theta$ and $0$ refer to non-zero and zero temperatures, respectively, $f_{1}$ is the Fermi distribution function for electrons with temperature $ \theta_{ }$ in lead $1$, which the additional electron is injected into, $G ^{(1)}_{1,0}$ is given in Eq.~(\ref{ap0301}), and the rapidity $c = \hbar/\left(  2 \tau_{ D} \Gamma _{\tau} \right)$, see $ \psi^{(c)}$, Eq.~(\ref{04}).  

The above equation can be interpreted as the correlation function for a single-particle mixed quantum state with component states distributed according to the thermal probability density $p( \epsilon) =   -  \partial f_{1} ( \epsilon) /  \partial \epsilon $  and having correlation functions $e^{i \frac{ \epsilon }{ \hbar } \tau } G ^{(1)}_{1, 0} \left(  t + \tau - \frac{ \epsilon }{ c  }; t - \frac{ \epsilon }{ c  } \right)$. 

\subsection{ The term with $- v_{ \mu}^{2} G_{1} ^{(1)}  {\cal G}_{0} ^{(1)*}$}

Using Eqs.~(\ref{t01}) and (\ref{cf01-1}), I calculate

\begin{eqnarray}
{\Pi}_{1}^{}(\omega) &=&  
\frac{ 1 }{ \pi  }
\int\limits_{- \infty}^{ \infty } d \tau 
e^{i \omega \tau }
\frac{ 1/  \tau_{ \theta_{ } }}{ \sinh\left(  \tau /  \tau_{ \theta_{ }  }    \right) }
\label{t02} \\
&&
\times
{\rm Im}
\int _{}^{ } d \epsilon e^{i \frac{ \epsilon }{ \hbar } }
\left(  - \frac{  \partial f_{1} }{  \partial \epsilon  } \right)
{\cal C}\left(  \tau \right) ,
\nonumber \\
{\cal C}\left(  \tau \right) &=& \int\limits _{ - \infty}^{ \infty } dt  e ^{ -i \frac{ \mu }{ \hbar } \tau}   v_{ \mu}  G ^{(1)}_{1, 0} \left(  t  - \frac{ \epsilon  }{ c  } + \tau; t - \frac{ \epsilon }{ c  } \right)
.
\nonumber  
\end{eqnarray}
\ \\ \noindent
Since energy $ \epsilon$ and time $t$ enter $G ^{(1)}_{1,0}$ as a difference,  $t - \epsilon/c$, the correlation function calculated at zero temperature and integrated over time, ${\cal C}\left(  \tau \right) $, becomes independent of $ \epsilon$. 
This property allows us to integrate out $ \epsilon$, 

\begin{eqnarray}
\Pi_{1}\left( \omega \right) &=&  
\int\limits _{- \infty}^{ \infty } d \tau e^{i \omega \tau} 
\left(  \frac{ \tau/  \tau_{ \theta_{ } }}{ \sinh\left(  \tau /  \tau_{ \theta_{ }  }    \right) } \right)^{2}
\frac{ {\rm Im}  \left[ {\cal C}\left(  \tau \right) \right] }{  \pi \tau  }
.
\quad
\label{t03}
\end{eqnarray}
\ \\ \noindent
With $G ^{(1)}_{1,0}$ from Eq.~(\ref{ap0301}), I calculate ${\cal C}\left(  \tau \right)$ as described above in Appendix~\ref{ap3s1} and find,

\begin{eqnarray}
{\Pi}_{1}^{}(\omega) =
\frac{ 1 }{ 2 \pi  \Gamma _{\tau} }
\int\limits_{- \infty}^{ \infty } 
d \tau  e^{i \omega \tau }
\left(  \frac{ \tau/  \tau_{ \theta_{ } }}{ \sinh\left(  \tau /  \tau_{ \theta_{ }  }    \right) } \right)^{2}
\frac{ 1
 }{1 +  \tau^{2}/\left( 2  \Gamma _{\tau} \right)^{2} }
. 
\nonumber \\
\label{t04} 
\end{eqnarray}
\ \\ \noindent
Remind that the thermal coherence time $ \tau_{ \theta_{ }} = \hbar /( \pi k_{B} \theta_{ 1})$. 
At $ \theta_{1} = 0$ we recover Eq.~(\ref{ap0305}). 
Note that ${\Pi}_{1}^{}(\omega)$  obeys Eq.~(\ref{ex04}). 

\subsection{ The term with  $ v_{ \mu}^{2} \left | G ^{(1)} \right |^{2}$}

At non-zero temperature the corresponding term reads,

\begin{eqnarray}
{\Pi}_{2}(\omega)  = 
\iint _{}^{ } d \epsilon d \epsilon ^{\prime}  
p( \epsilon) p( \epsilon ^{\prime})
\left |  {\cal N}_{\epsilon- \epsilon ^{\prime}}\left( \omega \right)  \right |^{2}  . 
\label{t05} 
\end{eqnarray}
\ \\ \noindent
Here the thermal probability density $p( \epsilon) =   -  \partial f_{1} ( \epsilon) /  \partial \epsilon $, and 

\begin{eqnarray}
{\cal N}_{\epsilon - \epsilon ^{\prime}}\left( \omega \right) = \frac{ 1 }{ \pi }
\int\limits_{- \infty}^{ \infty } d z e^{i \left(  \omega + \frac{  \epsilon - \epsilon ^{\prime} }{ \hbar }  \right)\Gamma _{\tau} z}  
\quad
\label{t06} \\
\int\limits_{0}^{ \infty} dx
e^{ - x } 
e^{   i x \left( z-  \frac{ \epsilon_{} }{ c \Gamma _{\tau}  }  \right)     } 
e^{ -i x^{2} \zeta ^{}  }
\int\limits_{0}^{ \infty} dy
e^{ - y } 
e^{  - i y \left( z-  \frac{ \epsilon_{} ^{\prime} }{ c \Gamma _{\tau}  }  \right)     } 
e^{ i y^{2} \zeta ^{}  } ,
\nonumber 
\end{eqnarray}
\ \\ \noindent
where $ z= \tau/  \Gamma _{\tau}$. 
Up to an irrelevant phase factor, ${\cal N}_{\epsilon- \epsilon ^{\prime}}\left( \omega \right)$ is the same as ${\cal N}\left( \omega \right)$ from Eq.~(\ref{ap0307}). 
Therefore, we can write, 

\begin{eqnarray}
\Pi_{2}\left(   \omega \right)  &=& 
\frac{ 1 }{1 + \left( \omega \tau_{D} \right)^{2}  }
\iint _{}^{ } d \epsilon d \epsilon ^{\prime}  p( \epsilon) p( \epsilon ^{\prime})
e^{-  \left |  \omega   + \frac{  \epsilon- \epsilon ^{\prime}  }{ \hbar } \right | 2 \Gamma _{\tau}}  .
\nonumber \\
\label{t07} 
\end{eqnarray}
\noindent \\
To bring the above equation into the form close to $ \Pi_{1}$, Eq.~(\ref{t04}), let us represent the exponential factor as follows,

$$
e^{-  \left |  \omega   + \frac{  \epsilon- \epsilon ^{\prime}  }{ \hbar } \right | 2 \Gamma _{\tau} } = 
\frac{ 1 }{ 2 \pi  \Gamma _{\tau} } 
\int\limits_{- \infty}^{ \infty } d \tau  
\frac{ e^{i \left(  \omega   + \frac{  \epsilon- \epsilon ^{\prime}  }{ \hbar } \right) \tau }
 }{1 +  \tau^{2}/\left( 2  \Gamma _{\tau} \right)^{2} } .
$$ 
Then we can integrate over $ \epsilon$ and $ \epsilon ^{\prime}$ in Eq.~(\ref{t07}) as follows,

$$
\int _{}^{ } d \epsilon p( \epsilon) e^{ i \frac{ \epsilon }{ \hbar  } \tau} = 
 \frac{ \tau/  \tau_{ \theta_{ } }}{ \sinh\left(  \tau /  \tau_{ \theta_{ }  }    \right) } ,
$$
and find that 

\begin{eqnarray}
\Pi_{2}\left(   \omega \right) &=& 
\frac{ 
1
}{ 1 + \left( \omega \tau_{D} \right)^{2}  } 
\int\limits_{- \infty}^{ \infty } \frac{d \tau e^{i \omega \tau } }{ 2 \pi  \Gamma _{\tau} } 
\frac{ \left(  \frac{ \tau/  \tau_{ \theta_{ } }}{ \sinh\left(  \tau /  \tau_{ \theta_{ }  }    \right) } \right)^{2}
 }{1 +  \tau^{2}/\left( 2  \Gamma _{\tau} \right)^{2} }
\nonumber \\ 
\label{t07-1} \\
&=& \frac{ \Pi_{1}\left(  \omega \right) }{ 1 + \left( \omega \tau_{D} \right)^{2} } ,
\nonumber 
\end{eqnarray}
\noindent \\ 
where $ \Pi_{1}$ is given in Eq.~(\ref{t04}).

Notice that we cannot use the inverse Fourier transformation with respect to $ \omega$ in order to express $\Pi_{2}\left(   \omega \right)$ at temperature $ \theta$ in terms of $\Pi_{2}\left(   \omega \right)$ at zero temperature. 
As a result, the excess noise, Eq.~(\ref{ap0309}), does not obey Eq.~(\ref{ex04}). 

\subsection{ Excess noise power}

The relation between $\Pi_{2}\left(   \omega \right)$ and $\Pi_{1}\left(   \omega \right)$, Eq.~(\ref{t07-1}), is independent on temperature. 
Therefore, in the case under consideration, temperature affects equally both auto- and cross-correlation noise at any frequency, 

\begin{eqnarray}
{\cal P}_{33}^{ex}(\omega) &=& 
R \frac{ e^{2} }{ {\cal T} _{0} }  
\frac{ T + \left( \omega  \tau_{D} \right)^{2}  }{ 1 + \left( \omega  \tau_{D} \right)^{2} } e^{-  \left | \omega \right | 2 \Gamma _{\tau}} 
 \eta_{ }\left(   \omega, \theta \right)
,
\nonumber \\
\label{t09} \\
{\cal P}_{34}^{ex}(\omega) &=& - 
R \frac{ e^{2} }{ {\cal T} _{0} }  
\frac{ T }{ 1 + \left( \omega  \tau_{D} \right)^{2} } e^{-  \left | \omega \right |  2 \Gamma _{\tau}} 
 \eta_{ }\left(   \omega, \theta \right)
,
\nonumber 
\end{eqnarray}
\ \\ \noindent
where

\begin{eqnarray}
 \eta_{ }\left(   \omega, \theta \right) = 
 e^{  \left | \omega \right | 2 \Gamma _{\tau} } 
 \int\limits_{- \infty}^{ \infty } \frac{d \tau e^{i \omega \tau } }{ 2 \pi  \Gamma _{\tau} } 
\frac{ \left(  \frac{ \tau/  \tau_{ \theta_{ } }}{ \sinh\left(  \tau /  \tau_{ \theta_{ }  }    \right) } \right)^{2}
 }{1 +  \tau^{2}/\left( 2  \Gamma _{\tau} \right)^{2} }
.
\label{t08}
\end{eqnarray}
\noindent \\
This factor is shown in Eq.~(\ref{c02}), where I introduced $x = \tau/ \left(  2 \Gamma _{\tau} \right)$.


\end{document}